\documentclass[english,reprint, prap, twocolumn, superscriptaddress, longbibliography]{revtex4-2}
\usepackage[T1]{fontenc}
\usepackage[utf8x]{inputenc}
\usepackage{geometry}
\usepackage{color}
\usepackage{babel}
\usepackage{units}
\usepackage{textcomp}
\usepackage{amsmath}
\usepackage{amssymb}
\usepackage{graphicx}
\usepackage{esint}
\usepackage[unicode=true,pdfusetitle,
 bookmarks=true,bookmarksnumbered=false,bookmarksopen=false,
 breaklinks=false,pdfborder={0 0 0},pdfborderstyle={},backref=false,colorlinks=true]
 {hyperref}



\begin{document}
\global\long\def\vect#1{\overrightarrow{\mathbf{#1}}}%

\global\long\def\abs#1{\left|#1\right|}%

\global\long\def\av#1{\left\langle #1\right\rangle }%

\global\long\def\ket#1{\left|#1\right\rangle }%

\global\long\def\bra#1{\left\langle #1\right|}%

\global\long\def\tensorproduct{\otimes}%

\global\long\def\braket#1#2{\left\langle #1\mid#2\right\rangle }%

\global\long\def\omv{\overrightarrow{\Omega}}%

\global\long\def\inf{\infty}%

\title{Environmental Sensitivity of Fabry-Perot Microcavities Induced By
Layered Graphene-Dielectric Hybrid Coatings }
\date{\today}
\author{Rui Peixoto}
\thanks{Contributed Equally to the Work}
\address{Centro de Física das Universidades do Minho e Porto and Faculty of
Sciences, University of Porto, 4169-007 Porto, Portugal}
\author{J. P. Santos Pires}
\email{up201201453@fc.up.pt}

\address{Centro de Física das Universidades do Minho e Porto and Faculty of
Sciences, University of Porto, 4169-007 Porto, Portugal}
\author{Catarina S. Monteiro}
\email{catarina.s.monteiro@inesctec.pt}

\address{Institute for Systems and Computer Engineering, Technology and Science
(INESC TEC) and Department of Physics and Astronomy, Faculty of Sciences,
University of Porto, Rua do Campo Alegre 687, 4169-007 Porto, Portugal}
\address{Faculty of Engineering, University of Porto, R. Dr. Roberto Frias,
4200-465 Porto, Portugal}
\author{Maria Raposo}
\address{Laboratory of Instrumentation, Biomedical Engineering and Radiation
Physics (LIBPhys-UNL), Department of Physics, NOVA School of Science
and Technology, NOVA University Lisbon, 2829-516 Caparica, Portugal}
\author{Paulo A. Ribeiro}
\address{Laboratory of Instrumentation, Biomedical Engineering and Radiation
Physics (LIBPhys-UNL), Department of Physics, NOVA School of Science
and Technology, NOVA University Lisbon, 2829-516 Caparica, Portugal}
\author{Susana O. Silva}
\address{Institute for Systems and Computer Engineering, Technology and Science
(INESC TEC) and Department of Physics and Astronomy, Faculty of Sciences,
University of Porto, Rua do Campo Alegre 687, 4169-007 Porto, Portugal}
\author{Orlando Frazão}
\address{Institute for Systems and Computer Engineering, Technology and Science
(INESC TEC) and Department of Physics and Astronomy, Faculty of Sciences,
University of Porto, Rua do Campo Alegre 687, 4169-007 Porto, Portugal}
\author{J. M. Viana Parente Lopes}
\email{jlopes@fc.up.pt}

\address{Centro de Física das Universidades do Minho e Porto and Faculty of
Sciences, University of Porto, 4169-007 Porto, Portugal}
\begin{abstract}
We propose a fiber-based environmental sensor that exploits the reflection
phase shift tunability provided by the use of layered coatings composed
of dielectric slabs spaced by conducting membranes.\,A transfer matrix
study is done in a simplified theoretical model, for which an enhanced
sensitivity of the reflection interference pattern to the output medium
is demonstrated, in the typical refractive index range of liquid media.
An experimental configuration using a cascaded Fabry-Perot microcavity
coated by a graphene oxide/polyethylenimine (GO/PEI) multilayered
structure is demonstrated. Its cost effective chemical production
method makes graphene oxide-based hybrid coatings excellent candidates
for future real-life sensing devices.
\end{abstract}
\maketitle

\section{Introduction}

\vspace{-0.2cm}

One of the main technological application of optical fibers is to
fabricate sensing devices that can detect the tiniest changes in physical
or chemical parameters. Such measurements can be very precise but
are usually indirect and exploit the effect of small parametric changes
in the propagation of guided lightwaves. The accuracy of these sensors
relies on interferometry that is made possible, for instance, by building
hollow core microcavities (as small as $30\text{\textmu m}$\,\citep{Lee2015,Gomes2019})
using capillary tubes or photonic crystal fibers (PCFs) at end of
a cleaved optical fiber\,\citep{Sirkis_1993,yu_2003,wang_2006,habel_2011,Hu2012_MuSpheres,favero2012_MuSpheres,Jauregui2013_MuSpheres,Zou_2013,Lee2015,ferreira_2015,monteiro_2016,Yan2017_MuSpheres,Gomes2018_MuSpheres,wang_2018,Monteiro2019_MuSpheres,monteiro_2021},
that act as miniature Fabry-Perot interferometers. Fabry-Perot cavities
in optical fibers were firstly fabricated by Sirkis \textit{et al.}
for strain measurement\,\citep{Sirkis_1993} and, since then, similar
interferometric sensors have been proposed for curvature\,\citep{monteiro_2016},
temperature\,\citep{ferreira_2015}, pressure\,\citep{wang_2006},
humidity\,\citep{wang_2018}, acoustic sensing\,\citep{monteiro_2021},
among others. Besides, fiber optic sensors based on Fabry-Perot interferometers
have also attracted much attention in other fields of expertise, ranging
from medicine\,\citep{Zou_2013} to engineering\,\citep{yu_2003,habel_2011}.

The periodicity of the interference fringes can be made dependent
on the refractive index of the environment\,($n_{\text{out}}$).\,This
can be achieved by endowing the output interface with a reflection
phase shift sensitive to the wavelength ($\lambda$), an effect that
can be produced by a thin coating which may feature an internal layered
structure built at the nanoscale\,\citep{baumeister2004}. Different
kinds of coatings have been used over the years, including dielectric
multilayered structures\,\citep{martin1997_Multilayer,jiang2009_Multilayer,Yehr2005},
porous metal-oxide films\,\citep{yoldas1980_Porous} and metallic
coatings based on thin metal films\,\citep{sung2002_metallic,chen2017_metallic},
plasmonic nanostructures\,\citep{kravets_plasmonic_2008} or embedded
nanoparticles\,\citep{krogman2005_nano,choi2014_nano}. More recently,
it has been reported\,\citep{kats2013} that hybrid coatings made
of lossy dielectrics (transparent narrow-gap semiconductors) deposited
on very thin highly conducting films can show nontrivial internal
interference effects that allow for an effective control over the
surface reflection coefficient of light at sub-wavelength scales.
With the advent of graphene\,\citep{Novoselov2004,Geim2007} and
other \textit{atomically thin planar conductors} (ATPC), it became
possible to realize similar ideas in hybrid optical coatings comprised
of dielectric-ATPC-dielectric multilayered structures. In effect,
if two-dimensional graphene precursors are employed, such hybrid coatings
can be produced by using layer-by-layer chemical methods\,\citep{keeney2015,Bjornmalm17}
that take advantage of intermolecular interactions between oppositely
charged electrolytes to self-assemble a multilayered structure\,\citep{kovtyukhova_1999}.
This inexpensive thin-film assembly already proved to be a viable
technique for real-life optical sensors of pH\,\citep{goicoechea_2008},
oxygen detection\,\citep{elosua_2015} and fluorescence\,\citep{Goncalves_2012}.

In this paper, we propose a fiber-based environmental sensor that
consists of a Fabry-Perot microcavity coated by a hybrid structure
made of dielectric polymer slabs separated by conducting two-dimensional
membranes. This sensor makes use of the aforementioned nontrivial
interference effects inside hybrid coatings, significantly enhancing
the sensitivity of the reflection interference fringes to the output
refractive index, in a range typical of liquid media. For concreteness,
our theoretical analysis considers the conducting membranes to be
pristine graphene sheets, whose optical conductivity is real and roughly
$\lambda$-independent in the mid- to near-infrared range\,\citep{Stauber2008,Nair2008,Peres2010}
(in the sensor's operation regime). However, the precise nature of
these membranes is insubstantial and the conclusions presented here
only require them to be two-dimensional materials with sufficiently
large real optical conductivities. Hence, more practical realizations
of these devices can use hybrid optical coatings based on graphene-oxide
(GO) sheets\,\citep{Hummers1958,Mkhoyan2009,schoche_optical_2017},
instead of actual graphene monolayers. As referred, layered structures
based on these graphene precursors can be grown by inexpensive chemical
processes in controlled layer-by-layer\,\citep{kovtyukhova_1999,Richardson16,Bjornmalm17}
procedures. Here, we employ this technique to experimentally realize
the proposed sensor in a cascaded Fabry-Perot interferometer at the
tip of a single-mode fiber that was coated by stacked polyethyleneimine/graphene-oxide
(PEI/GO) layers. Such devices have been recently proposed by some
of us\,\citep{Monteiro2020} although analyzed from a different viewpoint.

The paper is organized as follows: In Sec.\,\ref{sec:OpticalResponse},
we review the optical properties of pristine graphene and highlight
the similarities with existing studies on GO lattices. In Sec.\,\ref{sec:TransferMatrix},
we devise our theoretical model for the sensor, as well as the transfer
matrix formalism used to investigate the reflection properties of
the hybrid dielectric-ATPC-dielectric coatings, as a function of the
output refractive index and the system's parameters. We also present
the theory of interference in a low-finesse Fabry-Perot cavity, which
is a key ingredient to understand the physics behind the sensor's
operation. In Sec.\,\ref{sec:Resonance}, we present our main theoretical
predictions regarding the operation and performance of our idealized
environmental sensor. We also point out the chief advantages of using
graphene-based hybrid coatings over alternative setups. An experimental
realization of this sensing device is obtained in Sec.\,\ref{sec:RExperimental},
including measurements that demonstrate the operation predicted by
our theoretical analysis. Finally, in Sec.\,\ref{sec:Conclusions},
we sum up our conclusions and speculate on future applications of
this technology.

\vspace{-0.55cm}

\section{\label{sec:OpticalResponse}Optical Response of Graphene and Graphene-Oxide}

\vspace{-0.35cm}

The crucial ingredient of this sensor is the possibility to coat the
outward surface of the Fabry-Perot microcavity by a layered structure
of dielectric slabs interfaced by conducting two-dimensional membranes.
As explained in the introduction, the detailed nature of these membranes
is not important for our conclusions, which only depend on a dissipative
optical response, that is sufficiently strong but weakly dependent
on the wavelength. In other words, the membrane must behave as a two-dimensional
metal in the sensor's operational range (i.e., the mid- to near-infrared
spectral range). Nevertheless, the simplest albeit paradigmatic example
of such a material is monolayer graphene, whose optical response will
be concisely reviewed here. In addition, recent studies on the optical
response of (more complicated) GO will also be referred, highlighting
the similarities with graphene that justify a direct comparison between
theory and experiments is this paper.

\begin{figure}[t]
\includegraphics[scale=0.15]{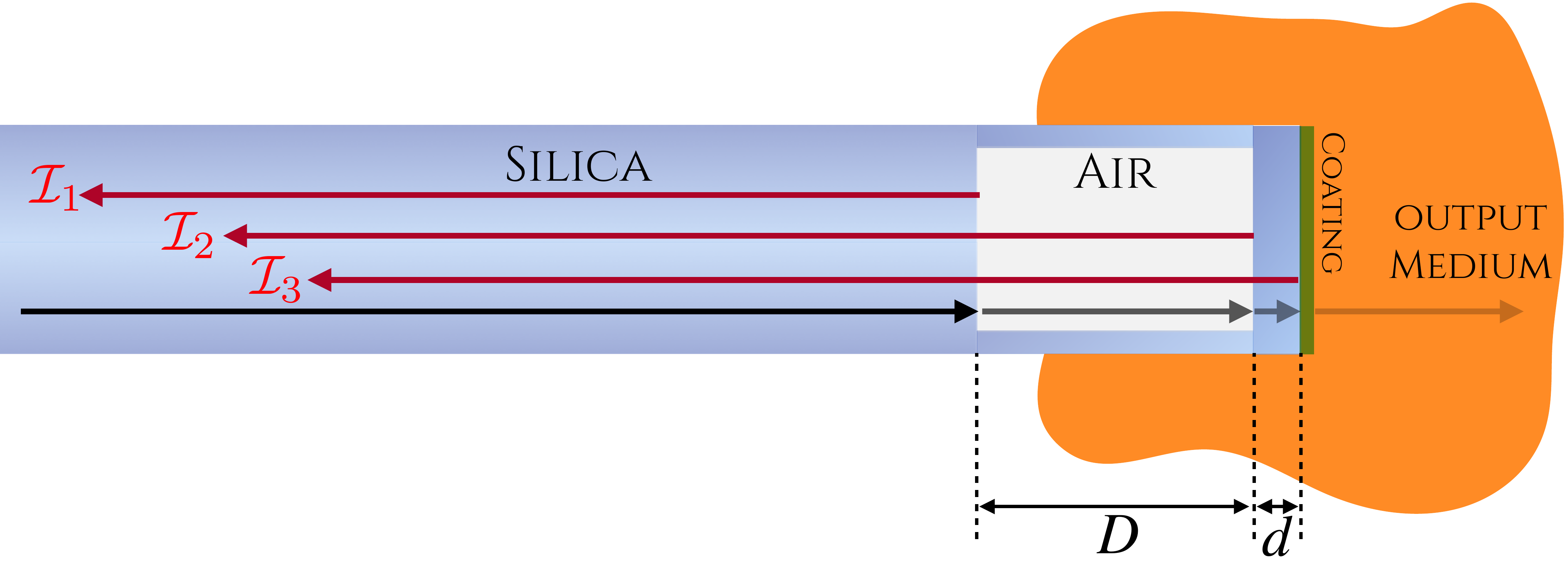}

\vspace{-0.2cm}

\caption{\label{fig:Scheme}Scheme of our fibre-based sensor using a three-wave
micro-metric Fabry-Perot cavity with an outside multilayered hybrid
surface coating (shown in green) made of conducting membranes spaced
by ideal dielectrics. We also define the dimensions of the Fabry-Perot
microcavity ($D$ and $d$) and depict the three primary waves reflected
off the interfaces of the cavity (of intensity $\mathcal{I}_{1}$,
$\mathcal{I}_{2}$ and $\mathcal{I}_{3}$). (color online)}

\vspace{-0.5cm}
\end{figure}

When working around the near- to mid-infrared regime ($\lambda\!\gtrsim\!800\text{nm}$),
it is known that the linear optical response of graphene is isotropic
($\sigma_{\lambda}^{xx}\!=\!\sigma_{\lambda}^{yy}\!=\!\sigma_{\lambda}$
are the only non-zero terms of the conductivity tensor) and well described
by continuum limit calculations done within a low-energy model having
two uncoupled Dirac-cones\,\citep{Stauber2008,Mikhailov2016,Ventura2017}.
More realistic modeling was also done\,\citep{cysne2016,Joao2019,KITE2020}
and confirm this approximation as appropriate. Under this approximation,
the low temperature optical conductivity of graphene can be shown
to have two contributions,

\vspace{-0.7cm}

\begin{equation}
\sigma_{\lambda}\!=\!\sigma_{\lambda}^{\text{intra}}\!+\!\sigma_{\lambda}^{\text{inter}},\label{eq:OpticalConductivityGraphene}
\end{equation}

\begin{figure}[t]
\begin{centering}
\hspace{-0.3cm}\includegraphics[scale=0.12]{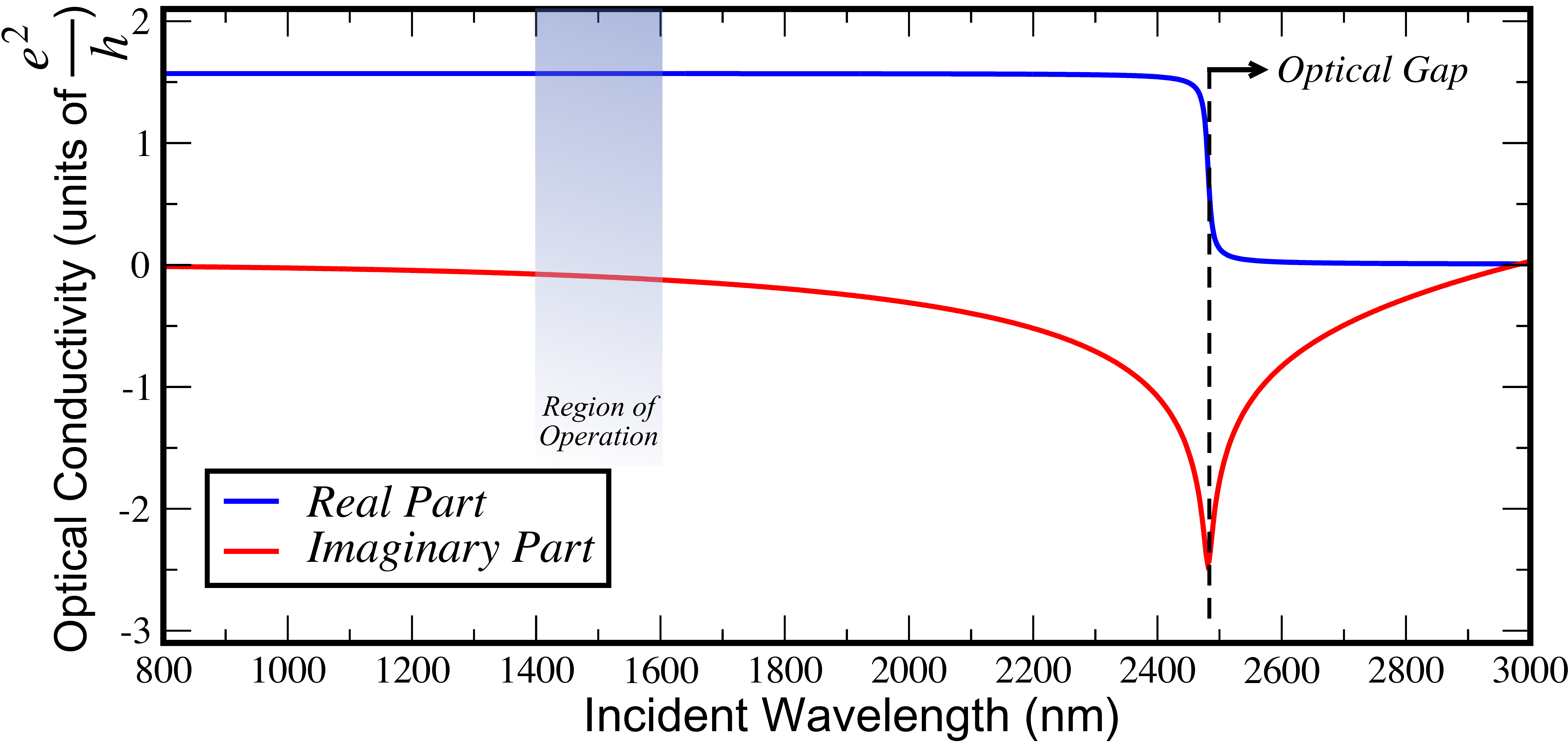}
\par\end{centering}
\vspace{-0.4cm}

\caption{\label{fig:OpticalConductivity} Plots of the complex optical conductivity
of a doped pristine graphene sheet in the infrared (see Eqs.\,(\ref{eq:Intraband})-(\ref{eq:Intraband-1})).
The plots consider the Fermi energy of $E_{\text{F}}\!=\!0.25eV$
and a relaxation scale of $\gamma\!=\!0.001eV$. The optical gap and
the operational regime for our sensor are indicated. (color online)}
\vspace{-0.5cm}
\end{figure}

\vspace{-0.2cm}

\noindent which stem from electronic intra- and inter-band transitions,
respectively. Both contributions can be analytically evaluated and
read\,\citep{Mikhailov2016},

\vspace{-0.3cm}

\begin{subequations}
\begin{equation}
\sigma_{\lambda}^{\text{intra}}\!=\!\frac{e^{2}}{h}\left(\frac{2\gamma\lambda^{2}\abs{E_{\text{F}}}}{h^{2}c^{2}\!+\!\gamma^{2}\!\lambda^{2}}\!+\!i\frac{2hc\lambda\abs{E_{\text{F}}}}{h^{2}c^{2}\!+\!\gamma^{2}\!\lambda^{2}}\right)\label{eq:Intraband}
\end{equation}

\vspace{-0.3cm}

\begin{align}
\sigma_{\lambda}^{\text{inter}}\! & =\!\frac{e^{2}}{2h}\arctan^{2}\left[2\abs{E_{\text{F}}}-\frac{h}{\lambda}\gamma\right]\label{eq:Intraband-1}\\
 & \qquad+i\frac{e^{2}}{4h}\ln\left[\frac{\gamma^{2}\lambda^{2}+\left(ch-2\abs{E_{\text{F}}}\lambda\right)^{2}}{\gamma^{2}\lambda^{2}+\left(ch+2\abs{E_{\text{F}}}\lambda\right)^{2}}\right],\nonumber 
\end{align}
\end{subequations}
 where $\gamma$ is a phenomenological scattering parameter, $c$
is the speed of light in the vaccum, $h$ is Planck's constant and
$E_{\text{F}}$ is the Fermi energy of graphene. For this approximate
calculations to hold, one must always consider that $\abs{E_{\text{F}}}\lesssim0.3\text{eV}$,
corresponding to a slightly doped graphene monolayer. In Fig.\,\ref{fig:OpticalConductivity},
we show some representative plots of these conductivities where it
is evident that $2\!\abs{E_{\text{F}}}$ acts as an effective ``optical
gap'' that marks a boundary between two response regimes. For $\lambda\!>\!\pi\hbar c/\!\abs{E_{\text{F}}}$,
the dissipative inter-band response is suppressed by Pauli-blocking
and a mostly reactive current appears on the system. In contrast,
for $\lambda\!<\!\pi\hbar c/\!\abs{E_{\text{F}}}$, the imaginary
part of the conductivity is almost zero and the response is mostly
due to inter-band transitions. Then, the optical response of graphene
becomes purely dissipative with a $\lambda$-independent universal
conductivity\,\citep{Nair2008}. The latter is the regime where our
proposed sensor will operate, such that we may take the optical conductivity
to be $\sigma_{\lambda}\!\approx\!e^{2}/4\hbar$.

As noted in the introduction, although pristine graphene could be
used to realize our proposal, cheaper and more practical alternatives
are provided by other two-dimensional materials, such as graphene-oxide
monolayers. These functionalized versions of graphene have considerably
more complex structures\,\citep{Mkhoyan2009} and simple expressions
cannot be readily obtained for their electronic or optical response
properties. However, due to its interest for optoelectronic applications\,\citep{Huang2020},
many \textit{Density Functional Theory} (DFT) studies have been published
on the linear optical properties of oxidized graphene sheets\,\citep{Huang2012,Nasehnia2015,Nasehnia2016,schoche_optical_2017}.
Most of the results demonstrate that, at least for moderately oxidized
graphene, the optical conductivity retains a purely dissipative plateau
in the mid-infrared\,\citep{Nasehnia2016} with a value approximating
the universal conductivity found in pristine graphene. Based on this,
we argue that the theory we will develop for coatings based off pristine
graphene membranes, also holds for GO, provided the value of the universal
optical conductivity is adjusted as a phenomenological parameter.

\vspace{-0.55cm}

\section{\label{sec:TransferMatrix}Theoretical model for the proposed sensor }

\vspace{-0.35cm}

We propose a fiber-based interferometric sensor (see Fig.\,\ref{fig:Scheme}a)
that operates on the basis of a reflection interference pattern that
is sensitive to its environment.\,Any changes in physical or chemical
properties of this surrounding medium typically translate into slight
alterations in its refractive index.\,For a bare output interface,
the sensor would only be affected by variations in the absolute reflectivity
of the last interface, as the phase introduced upon reflection is
independent\,\footnote{In these conditions, it can only change from $0$ to $\pi$ if the
output refractive index ($n_{\text{out}}$) exceeds that of silica.} of the incident wavelength.\,This implies that the interference
pattern in the reflected spectrum will be affected solely in the relative
strength of its Fourier components, but not in their periods.\,To
allow a shift in the fringe periods by environmental changes, one
must engineer the output interface so as to give it a $\lambda$-dependent
complex reflection coefficient that also depends parametrically on
the refractive index of the output medium. Such properties can be
embedded into the optical interface by deposing a nano-structured
coating onto the cavity's outwards surface, as described in the introduction.

Here, we present the basic theoretical model employed to simulate
the operation of our device proposal. Before proceeding with details,
it is important to highlight that the sensor is naturally split into
two main parts: i) the hybrid coating on the output interface and
ii) the Fabry-Perot microcavity at fiber's termination. Both components
work together to produce the interference effects on which the sensor
is based, however they act at different length scales. The hybrid
coating is structured at the nanoscale, with typical distances of
$\approx\!10\text{nm}$ between successive conducting membranes. These
lengths are shorter than optical wavelengths and thus responsible
for the internal interference that causes nontrivial phase shifts
upon reflection. In addition, one also has the optical interfaces
forming the Fabry-Perot microcavity. These surfaces are distanced
by tens/hundreds of \textmu m, scales that far exceed the ones present
in the internals of the hybrid coating. The purpose of this cavity
is simply to generate a basic three-wave interference pattern in the
reflected spectrum which will then be shaped by variations in the
output refractive index. In the following, we shall consider these
two main parts separately. 

\vspace{-0.6cm}

\subsection{\label{subsec:Optical-Modeling-of}Optical Modeling of a Hybrid Nano-structured
Coating}

\vspace{-0.3cm}

\begin{figure}[t]
\vspace{-0.4cm}
\begin{centering}
\hspace{-0.1cm}\includegraphics[scale=0.165]{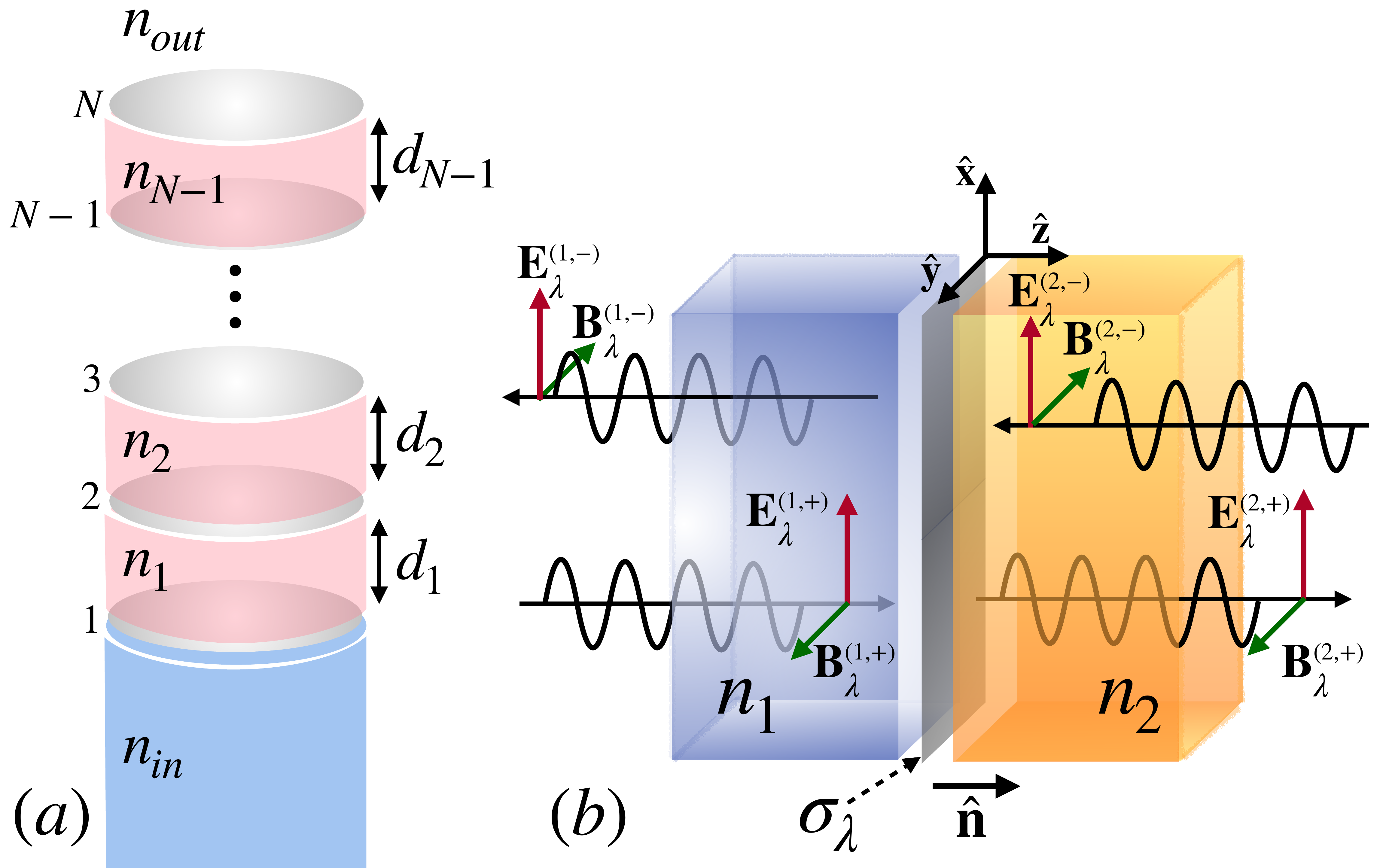}
\par\end{centering}
\vspace{-0.2cm}

\caption{\label{fig:Schemes}\textit{(a) }Toy-model used to investigate the
optical coefficients of layered hybrid coating. For concreteness,
we have considered the conducting membranes as doped pristine graphene
monolayers operating way above the optical gap. \textit{(b) }Scheme
of a perpendicular scattering setup for a single graphene interface
between two bulk dielectrics. (color online)}

\vspace{-0.5cm}
\end{figure}

The layer-by-layer assembly method creates optical coatings where
several (roughly parallel) conducting membranes are separated by nanoscopic
slabs of identical dielectric material. In practice, the control over
orientation, flatness and distances between sheets is not perfect
but, for a modeling purpose, we shall consider the conducting membranes
as mutually parallel planes as shown in Fig.\,\ref{fig:Schemes}a.\,Moreover,
since our proposal is based on a single-mode optical fiber, it is
appropriate to consider that the guided light-wave has normal incidence
on the coating, which simplifies the following calculations.\,In
short, our aim will be to evaluate the complex reflection/transmission
coefficients of a hybrid coating featuring a given number ($N$) of
conducting membranes spaced by a set of dielectric slabs characterized
by the widths, $\{d_{1},\!\cdots\!,d_{N\!-\!1}\!\}$. This problem
can be neatly solved by a \textit{Transfer Matrix Method} (TMM). With
no loss of generality, we consider the membranes to be pristine graphene
monolayers working at the universal optical conductivity regime,\,i.e.\,$\lambda\!\approx\!1500\text{nm}$,
while the dielectric slabs will be taken as made up of the same material
(of refractive index $n_{\text{pol}}$).

To devise a TMM that evaluates the complex transmission ($t_{\!\text{\ensuremath{\lambda}}}$)
and reflection coefficients ($r_{\!\text{\ensuremath{\lambda}}}$)
of the entire coating, we begin by considering a monochromatic wave
at normal incidence onto a single conducting plane (depicted in Fig.\,\ref{fig:Schemes}b).\,In
this case, both the electric and magnetic fields are parallel to the
graphene sheet and the solution of the four-wave scattering problem
boils down to imposing appropriate boundary conditions at the interface.
Since the interface is conducting, it supports surface currents driven
by the electric field of the crossing wave which, in turn, alter the
usual boundary conditions associated to a purely dielectric interface.
The general electromagnetic boundary conditions in the presence of
a surface conductivity are derived in the Appendix\,\ref{BoundaryConditions}.
Meanwhile, the electric field of the partial waves in Fig.\,\ref{fig:Schemes}b
can be written as

\vspace{-0.5cm}
\begin{equation}
\!\!\mathbf{E}^{1}\!\left(\mathbf{r},t\right)\!=\!\!\left(\!E_{\lambda}^{{\scriptscriptstyle (\!1,\!+\!)}}\!e^{i\frac{2\pi n_{1}z}{\lambda}}\!+\!E_{\lambda}^{{\scriptscriptstyle (\!1,\!-\!)}}\!e^{-i\frac{2\pi n_{1}z}{\lambda}}\!\right)\!e^{i\frac{2\pi c}{\lambda}t}\,\hat{\mathbf{x}}\label{eq:ElectricField1}
\end{equation}

\vspace{-0.2cm}

\noindent in the input medium (refractive index $n_{1}$) and

\vspace{-0.5cm}

\begin{equation}
\!\!\mathbf{E}^{2}\!\left(\mathbf{r},t\right)\!=\!\!\left(\!E_{\lambda}^{{\scriptscriptstyle (2,\!+\!)}}\!e^{i\frac{2\pi n_{2}z}{\lambda}}\!\!+\!E_{\lambda}^{{\scriptscriptstyle (2,\!-\!)}}\!e^{-i\frac{2\pi n_{2}z}{\lambda}}\!\right)\!e^{i\frac{2\pi c}{\lambda}t}\,\hat{\mathbf{x}}\label{eq:ElectricField2}
\end{equation}

\vspace{-0.2cm}

\noindent in the output one (refractive index $n_{2}$).\,The magnetic
components of the wave are also harmonic and can be readily obtained
from Faraday's law --- $\mathbf{B}^{{\scriptscriptstyle (\pm)}}\!\!=\!\!\pm\hat{\mathbf{z}}\!\times\!\mathbf{E}^{{\scriptscriptstyle (\pm)}}\!/v_{m}$---
where $v_{m}\!=\!c/n_{m}$ is the speed of light in the corresponding
medium. These magnetic fields read

\vspace{-0.6cm}

\begin{equation}
\mathbf{B}^{1}\!\!\left(\mathbf{r},\!t\right)\!=\!\!\left(\!E_{\lambda}^{{\scriptscriptstyle (\!1,\!+\!)}}\!e^{i\frac{2\pi n_{1}z}{\lambda}}\!\!\!\!-\!\!E_{\lambda}^{{\scriptscriptstyle (\!1,\!-\!)}}\!e^{-i\frac{2\pi n_{1}z}{\lambda}}\!\right)\!\!e^{i\frac{2\pi ct}{\lambda}}\!\frac{c\,\hat{\mathbf{y}}}{n_{1}}\label{eq:MagneticField1}
\end{equation}

\vspace{-0.8cm}

\begin{equation}
\mathbf{B}^{2}\!\!\left(\mathbf{r},\!t\right)\!=\!\!\left(\!E_{\lambda}^{{\scriptscriptstyle (\!2,\!+\!)}}\!e^{i\frac{2\pi n_{2}z}{\lambda}}\!\!\!\!-\!\!E_{\lambda}^{{\scriptscriptstyle (\!2,\!-\!)}}\!e^{-i\frac{2\pi n_{2}z}{\lambda}}\!\right)\!\!e^{i\frac{2\pi ct}{\lambda}}\!\frac{c\,\hat{\mathbf{y}}}{n_{2}},\label{eq:MagneticField2}
\end{equation}

\vspace{-0.2cm}

\noindent with their orientation depicted also in Fig.\,(\ref{fig:Schemes})b.

Since the wave is at normal incidence, all the fields involved are
parallel to the conducting interface. This leads to two important
simplifications in the boundary conditions at $z\!=\!0$: i) Only
the conditions on field components parallel to the surface are of
importance; and ii) Despite generating surface currents, the time-varying
electric fields do not induce any superficial charge density waves.
In this case, the boundary conditions of the problem reduce to the
following simpler equations:

\vspace{-0.6cm}

\begin{subequations}
\begin{equation}
\mathbf{E}_{\lambda}^{1}=\mathbf{E}_{\lambda}^{2}\label{eq:ElectricFieldContinuity}
\end{equation}

\vspace{-0.8cm}

\begin{equation}
\mathbf{B}_{\lambda}^{1}\!=\!\mathbf{B}_{\lambda}^{2}\!+\!\mu_{0}\sigma_{\lambda}\mathbf{\hat{z}}\!\times\!\mathbf{E}_{\lambda}^{2},\label{eq:MagneticFieldCondition}
\end{equation}
\end{subequations}

\vspace{-0.25cm}

\noindent where $\mathbf{E}_{{\scriptscriptstyle 1/2}}$ and $\mathbf{B}_{{\scriptscriptstyle 1/2}}$
are the monochromatic vector amplitudes defined in Eqs.\,(\ref{eq:ElectricField1})-(\ref{eq:MagneticField2}),
$\sigma_{\lambda}$ is the optical surface conductivity of the planar
membrane and $\mu_{0}$ is the vaccum magnetic permeability\,\footnote{We will always assume the magnetic effects to be irrelevant for the
optical properties of our system.}.\,By expressing Eqs.\,(\ref{eq:ElectricFieldContinuity}) and (\ref{eq:MagneticFieldCondition})
in terms of only incoming/outgoing electric field amplitudes, $E_{\lambda}^{{\scriptscriptstyle (\!1/2,\!\pm\!)}}$,
we cast the scattering problem in the matrix form,

\vspace{-0.55cm}

\begin{equation}
\left(\!\!\begin{array}{c}
E_{\lambda}^{{\scriptscriptstyle (\!2,\!+\!)}}\\
E_{\lambda}^{{\scriptscriptstyle (\!2,\!-\!)}}
\end{array}\!\!\right)\!=\!\mathbb{T}_{1,2}^{\text{G}}\!\left(\lambda\right)\cdot\left(\!\!\begin{array}{c}
E_{\lambda}^{{\scriptscriptstyle (\!1,\!+\!)}}\\
E_{\lambda}^{{\scriptscriptstyle (\!1,\!-\!)}}
\end{array}\!\!\right),\label{GrapheneInterfaceLinearRelation}
\end{equation}

\vspace{-0.2cm}

\noindent where the transfer matrix of a single conducting interface
between dielectrics of refractive indices $n_{1}$ and $n_{2}$ is
given by

\vspace{-0.5cm}

\begin{equation}
\mathbb{T}_{{\scriptscriptstyle 1,2}}^{\text{G}}\!(\lambda)\!=\!\!\left(\!\begin{array}{cc}
\!\frac{n_{2}+n_{1}}{2n_{2}}\!-\!\frac{\mu_{0}c\sigma_{\lambda}}{2n_{2}} & \!\frac{n_{2}-n_{1}}{2n_{2}}\!-\!\frac{\mu_{0}c\sigma_{\lambda}}{2n_{2}}\\
\!\frac{n_{2}-n_{1}}{2n_{2}}\!+\!\frac{\mu_{0}c\sigma_{\lambda}}{2n_{2}} & \!\frac{n_{2}+n_{1}}{2n_{2}}\!+\!\frac{\mu_{0}c\sigma_{\lambda}}{2n_{2}}
\end{array}\!\!\right).\label{eq:GrapheneTM}
\end{equation}

\vspace{-0.2cm}

Having the linear relation of Eq.\,(\ref{GrapheneInterfaceLinearRelation}),
it is trivial to generalize the result and relate the input to the
output field components in the whole structure of Fig.\,\ref{fig:Schemes}b.\,For
that, we recognize that a complex phase factor\,---\,$\exp\left(\pm i2\pi n_{i}d_{i}/\lambda\right)$\,---\,is
accumulated by right- and left-traveling components, respectively,
when traversing the dielectric slab between layers $i$ and $i\!+\!1$.\,Hence,
the electric field amplitudes at the input and output media can be
related as

\vspace{-0.5cm}
\begin{equation}
\left(\!\!\begin{array}{c}
E_{\lambda}^{{\scriptscriptstyle (\!in,\!+\!)}}\\
E_{\lambda}^{{\scriptscriptstyle (\!in,\!-\!)}}
\end{array}\!\!\right)\!=\!\mathcal{T}_{\lambda}\!\cdot\!\left(\!\!\begin{array}{c}
E_{\lambda}^{{\scriptscriptstyle (\!out,\!+\!)}}\\
E_{\lambda}^{{\scriptscriptstyle (\!out,\!-\!)}}
\end{array}\!\!\right),\label{eq:FullTransferMatrix1}
\end{equation}

\vspace{-0.2cm}

\noindent where $\mathcal{T}_{\lambda}$ is a $2\!\times\!2$ transfer
matrix characterizing the transmission of monochromatic electromagnetic
waves across the entire coating. This matrix is defined in terms of
the interface matrices\,---\,$\mathbb{T}_{{\scriptscriptstyle i,i+1}}^{\text{G}}\!(\lambda)$\,---\,defined
in Eq.\,(\ref{eq:GrapheneTM}), and the matrices\,---\, $\mathbb{P}_{i}\!=\!\text{diag}\left(\exp\left(i2\pi n_{i}d_{i}/\lambda\right),\exp\left(-i2\pi n_{i}d_{i}/\lambda\right)\right)$\,---\,
describing the phase-accumulation in each dielectric slab. The general
expression then reads,

\vspace{-0.45cm}

\begin{equation}
\mathcal{T}_{\lambda}\!\!=\!\mathbb{T}_{{\scriptscriptstyle 0,1}}^{\text{G}}\!(\lambda)\!\cdot\!\mathbb{P}_{1}\!\cdot\!\mathbb{T}_{{\scriptscriptstyle 1,2}}^{\text{G}}\!(\lambda)\!\cdot\,\cdots\,\cdot\!\mathbb{P}_{N-1}\!\cdot\!\mathbb{T}_{{\scriptscriptstyle N-1,N}}^{\text{G}}\!(\lambda),
\end{equation}

\vspace{-0.15cm}

\noindent where $n_{0}\!=\!n_{\text{in}}$ and $n_{N}\!=\!n_{\text{out}}$. 

Before moving on, it is worthy to remark that the TMM is suitable
for extremely efficient numerical evaluation, involving only successive
product of very small ($2\!\times\!2$) matrices. In our case, by
specifying $n_{\text{in}}$, $n_{\text{out}}$, the list of dielectric
widths $\left\{ d_{1},\cdots,d_{N-1}\right\} $ and the corresponding
refractive indices $\left\{ n_{1},\cdots,n_{N-1}\right\} $, the transfer
matrix of the whole coating can be computed in a matter of seconds
in the average laptop. The complex reflection and transmission coefficients
can then be extracted from its matrix elements:

\vspace{-0.55cm}

\begin{subequations}
\begin{equation}
r_{\lambda}\!=\!\frac{E_{\lambda}^{{\scriptscriptstyle (\!in,\!-\!)}}}{E_{\lambda}^{{\scriptscriptstyle (\!in,\!+\!)}}}\!=\!-\frac{\left[\mathcal{T}_{\lambda}\right]_{12}}{\left[\mathcal{T}_{\lambda}\right]_{22}}\label{eq:ReflexionCoeff}
\end{equation}

\vspace{-0.6cm}

\begin{equation}
t_{\lambda}\!=\!\frac{E_{\lambda}^{{\scriptscriptstyle (\!out,\!+\!)}}}{E_{\lambda}^{{\scriptscriptstyle (\!in,\!+\!)}}}\!=\!\left[\mathcal{T}_{\lambda}\right]_{11}\!\!-\!\frac{\left[\mathcal{T}_{\lambda}\right]_{12}\left[\mathcal{T}_{\lambda}\right]_{21}}{\left[\mathcal{T}_{\lambda}\right]_{22}}.\label{eq:TransmissionCoeff}
\end{equation}

\vspace{-0.2cm}
\end{subequations}

Recapping, the TMM provides a convenient and fast method to evaluate
the reflection coefficient of an hybrid coating, having any number
of conducting membranes with given interlayer separations, and functioning
at a given wavelength. Hence, in what follows we may take $r_{\lambda}$
as a known quantity, which one can readily evaluate for a specific
case, and that fully characterizes the reflection properties of the
output coated surface.

\vspace{-0.5cm}

\subsection{Reflection Interference Pattern in a Three-Wave Fabry-Perot Cavity }

\vspace{-0.25cm}

\begin{figure}[t]
\begin{centering}
\hspace{-0.3cm}\includegraphics[scale=0.135]{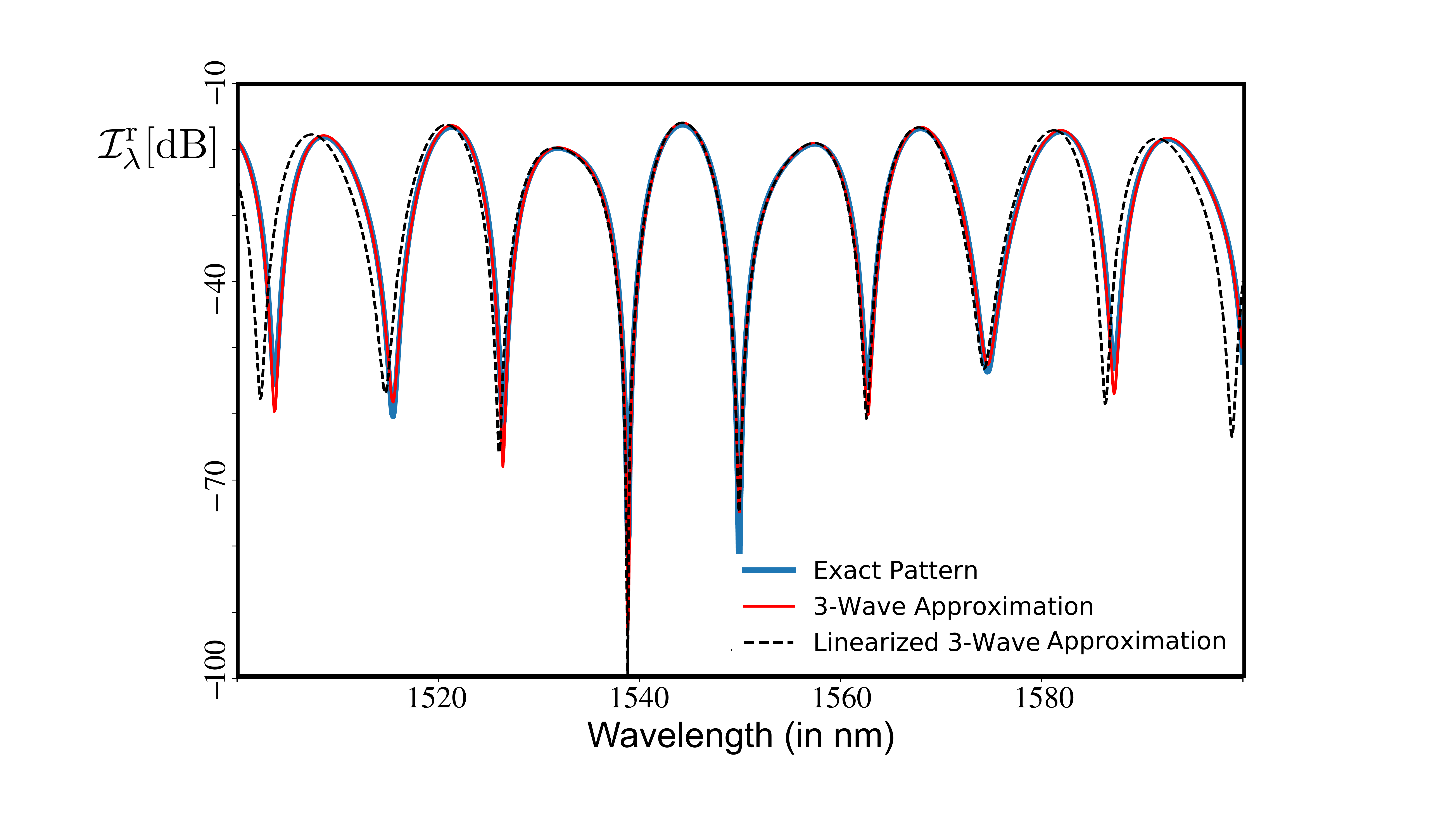}
\par\end{centering}
\vspace{-0.3cm}

\caption{\label{fig:MicrosphereIntereferometer}Intensity reflected by the
sensor as a function of the incident wavelength. The spectrum was
obtained for a \textit{simulated microcavity} ($D\!=\!100\mu\text{m}$
and $d\!=\!40\mu\text{m}$) coated by $N\!=\!20$ graphene monolayers
spaced by $20\text{nm}$ dielectric slabs ($n_{\text{pol}}\!=\!1.56$).
In blue, we have the exact interference pattern, obtained from a transfer-matrix
calculation on to the whole system (cavity + coating), while in red
and black dashed we show the three-wave approximation (see Eq.\,(\ref{eq:InterferencePattern_3Waves}))
and its linearized version (see Eq.\,(\ref{eq:InterferencePattern_3Waves-1}))
around $\lambda_{c}\!=\!1550\text{nm}$, respectively. (color online)}
\vspace{-0.6cm}
\end{figure}
 Besides the sub-wavelength processes within the hybrid coating, there
is also interference happening at larger scales, between the waves
reflected off the three interfaces that make up the low-finesse Fabry-Perot
microcavity at the fiber's termination (see cartoon in Fig.\,\ref{fig:Scheme}).\,Dealing
with this \textmu m-scale interference is not fundamentally different
from our previous analysis.\,However, in the interest of earning
a greater insight on the sensor's operation, we approach the problem
from a different angle and approximate the reflected signal by the
sum of the three primary reflections (the ones shown in Fig.\,\ref{fig:Scheme}).
This approximation is justified by the low reflectivity of all the
surfaces in the cavity and further confirmed by a direct comparison
with the reflected spectra obtained by the TMM applied to the whole
system (see Fig.\,\ref{fig:MicrosphereIntereferometer}). Considering
a three-wave approximation, the intensity reflected by the microcavity
is then given as

\vspace{-0.4cm}

\begin{align}
\mathcal{I}_{r}^{\lambda}\! & =\!\mathcal{I}_{1}\!+\!\mathcal{I}_{2}\!+\!\mathcal{I}_{3}^{\lambda}\!-2\sqrt{\mathcal{I}_{1}\mathcal{I}_{2}}\cos\!\left(\!\frac{4\pi D}{\lambda}\!\right)\label{eq:InterferencePattern_3Waves}\\
 & \qquad\qquad-2\sqrt{\mathcal{I}_{2}\mathcal{I}_{3}^{\lambda}}\cos\!\left(\!\frac{4\pi n_{s}d}{\lambda}\!+\!\theta_{c}^{\lambda}\!\right)\nonumber \\
 & \qquad\qquad+2\sqrt{\mathcal{I}_{1}\mathcal{I}_{3}^{\lambda}}\cos\!\left(\!\frac{4\pi(D\!+\!n_{s}d)}{\lambda}\!+\!\theta_{c}^{\lambda}\!\!\right),\nonumber 
\end{align}

\vspace{-0.2cm}

\noindent where $n_{s}$ is the refractive index of the silica, $d$
and $D$ are the cavity widths (as defined in Fig.\,\ref{fig:Scheme}),
and $\mathcal{I}_{1,2,3}$ are the intensities reflected by each interface.
These intensities can be obtained from Fresnel laws at normal incidence,
reading

\vspace{-0.5cm}

\begin{align}
\sqrt{\mathcal{I}_{1}} & \!=\!\abs{\frac{n_{s}\!-\!1}{n_{s}\!+\!1}}\!,\ \sqrt{\mathcal{I}_{2}}\!=\!\frac{4n_{s}\abs{n_{s}\!-\!1}}{\left(n_{s}\!+\!1\right)^{3}}\\
 & \qquad\qquad\qquad\qquad\,\text{and }\sqrt{\mathcal{I}_{3}}\!=\!\frac{16n_{s}^{2}R_{c}^{\lambda}}{\left(n_{s}\!+\!1\right)^{4}}.\nonumber 
\end{align}
As anticipated, effects from the output coating in the interference
pattern are encapsulated in its complex reflection coefficient, specified
by the corresponding reflectivity ($R_{\!c}^{\lambda}$) and reflection
phase shift ($\theta_{c}^{\lambda}$) in all previous expressions.
These parameters are crucially dependent on the incident wavelength
($\lambda$) and can be calculated by the TMM described in Subsect.\,\ref{sec:TransferMatrix}.
As we shall see, the $\lambda$-dependence of the reflection phase
shift plays a pivotal role in the operation of our sensor and is,
in effect, the central concept of this paper.

Regardless of coating details, one can point out some generic features
of the interference fringes. As evident from Eq.\,(\ref{eq:InterferencePattern_3Waves}),
even within a three-wave approximation, a Fabry-Perot interferometer
does not yield a periodic interference pattern upon reflection. Nevertheless,
if only a small neighborhood of a central wavelength ($\lambda_{\text{c}}$)
is observed, one can consider a lowest-order expansion around $\lambda_{c}$
and a three-component periodic pattern emerges, 

\vspace{-0.4cm}

\begin{align}
\mathcal{I}_{r}^{{\scriptscriptstyle \lambda_{\text{c}}\!+\!\Delta\lambda}} & \approx\mathcal{I}_{1}+\mathcal{I}_{2}+\!\mathcal{I}_{3}^{\lambda_{\text{c}}}\!\label{eq:InterferencePattern_3Waves-1}\\
- & 2\sqrt{\mathcal{I}_{1}\!\mathcal{I}_{2}}\cos\!\left(\!\frac{2\pi\Delta\lambda}{\Lambda_{1}}\!-\!\frac{4\pi D}{\lambda_{\text{c}}^{2}}\!\right)\!\nonumber \\
- & 2\sqrt{\mathcal{I}_{2}\mathcal{I}_{3}^{\lambda_{\text{c}}}}\cos\!\left(\!\frac{2\pi\Delta\lambda}{\Lambda_{2}}\!-\!\frac{4\pi n_{\text{s}}d}{\lambda_{\text{c}}^{2}}\!-\!\theta_{c}^{\lambda_{\text{c}}}\!\right)\nonumber \\
+ & 2\sqrt{\mathcal{I}_{1}\mathcal{I}_{3}^{\lambda_{\text{c}}}}\cos\!\left(\!\frac{2\pi\Delta\lambda}{\Lambda_{3}}\!-\!\frac{4\pi\left(D\!+\!n_{\text{s}}d\right)}{\lambda_{\text{c}}^{2}}\!-\!\theta_{c}^{\lambda_{\text{c}}}\!\right).\nonumber 
\end{align}

\noindent \vspace{-0.2cm}

\noindent This pattern features three effective periods ($\Lambda_{1,2,3}$)
that arise from the interference of any combination of the three partial
waves of Fig.\,\ref{fig:Scheme}. The three periods are then given
as:

\noindent \vspace{-0.6cm}
\begin{align}
\Lambda_{1}\!= & \frac{\lambda_{\text{c}}^{2}}{2D}\,,\,\Lambda_{2}\!=\frac{2\pi\lambda_{\text{c}}^{2}}{4\pi n_{\text{s}}d\!+\!\lambda_{\text{c}}^{2}\left.\nicefrac{d\theta_{c}^{\lambda}}{d\lambda}\right|_{\lambda_{\text{c}}}}\,\label{eq:lambda_eff}\\
 & \qquad\text{and}\,\Lambda_{3}\!=\frac{2\pi\lambda_{\text{c}}^{2}}{4\pi\left(D\!+\!n_{\text{s}}d\right)\!+\!\lambda_{\text{c}}^{2}\,\left.\nicefrac{d\theta_{c}^{\lambda}}{d\lambda}\right|_{\lambda_{\text{c}}}}.\nonumber 
\end{align}

\noindent \vspace{-0.3cm}

\noindent With no surprise, Eqs.\,(\ref{eq:lambda_eff}) demonstrate
that the effect of coating the output interface is a slight change
in $\Lambda_{2}$ and $\Lambda_{3}$, relative to their bare values.
These are the components of the pattern which arise from interference
with the partial wave reflected off the output interface. Interestingly,
they also show that the shift in the periods is due to a $\lambda$-dependent
reflection phase shift, something which would be absent in a bare
dielectric-to-dielectric interface. These shifts may increase (positive
response) or a decrease (negative response) the periods relative to
their bare values, depending on the sign of $\left.\nicefrac{d\theta_{c}^{\lambda}}{d\lambda}\right|_{\lambda_{\text{c}}}$
. As we shall see, both the sign and magnitude of this derivative
can be tuned by changes in the refractive index of the output medium,
thus providing the operational principle behind the cavity's sensitivity
to the environment. Note that, in practice, the individual periods
of the interference pattern can be studied by band-filtering the Fourier
transform of the reflected signal within a narrow enough spectral
window\,\citep{Monteiro2020}. 

\vspace{-0.5cm}

\section{\label{sec:Resonance}Nontrivial Reflection Phase Shifts and Enhanced
Response to Liquid Environments}

\vspace{-0.3cm} 

The dependence of the reflected interference fringes on the output
medium is determined by the phase shifts introduced when the wave
is reflected off the last interface of the microcavity. More precisely,
Eqs.\,(\ref{eq:lambda_eff}) show that a $\lambda$-dependent phase
shift changes the periods in the interference pattern, with respect
to their bare values (determined by the microcavity's geometry). If
the tip of the sensor was a bare dielectric-to-dielectric interface,
then the reflection dephasing would be independent of both $\lambda$
and $n_{\text{out}}$. In this case, any changes to the periodicity
of the fringes could only arise by geometrical alterations of the
microcavity itself. However, such situation does not hold if a thin
dielectric coating (of index $n_{\text{pol}}$ and width $\Delta$)
is placed on top of the output surface. Then, interference within
the coating can bring about nontrivial phase shifts on the last partial
wave, that usually depend on the wavelength. A straightforward transfer
matrix analysis of this simple setup reveals the following complex
reflection coefficient:

\vspace{-0.5cm}

\begin{equation}
r_{\!\text{c}}^{{\scriptscriptstyle \lambda}}\!\!=\!\frac{(\frac{n_{\text{s}}}{n_{\!\text{pol}}}\!\!-\!\!1)(1\!\!+\!\frac{n_{\!\text{out}}}{n_{\!{\scriptscriptstyle \text{pol}}}})\!+\!(1\!\!-\!\frac{n_{\!\text{out}}}{n_{\!\text{pol}}})(1\!\!+\!\!\frac{n_{\text{s}}}{n_{\!\text{pol}}})e^{\frac{4i\pi n_{\!\text{pol}}\!\Delta}{\lambda}}}{(\frac{n_{\text{s}}}{n_{\!\text{pol}}}\!\!+\!\!1)(1\!\!+\!\frac{n_{\!\text{out}}}{n_{\!\text{pol}}})\!-\!(1\!\!-\!\frac{n_{\!\text{out}}}{n_{\!\text{pol}}})(1\!\!-\!\!\frac{n_{\text{s}}}{n_{\!\text{pol}}})e^{\frac{4i\pi n_{\!\text{pol}}\!\Delta}{\lambda}}}.\label{eq:ReflectionCoeff}
\end{equation}

\begin{figure}[t]
\vspace{-0.4cm}

\hspace{-0.25cm}\includegraphics[scale=0.205]{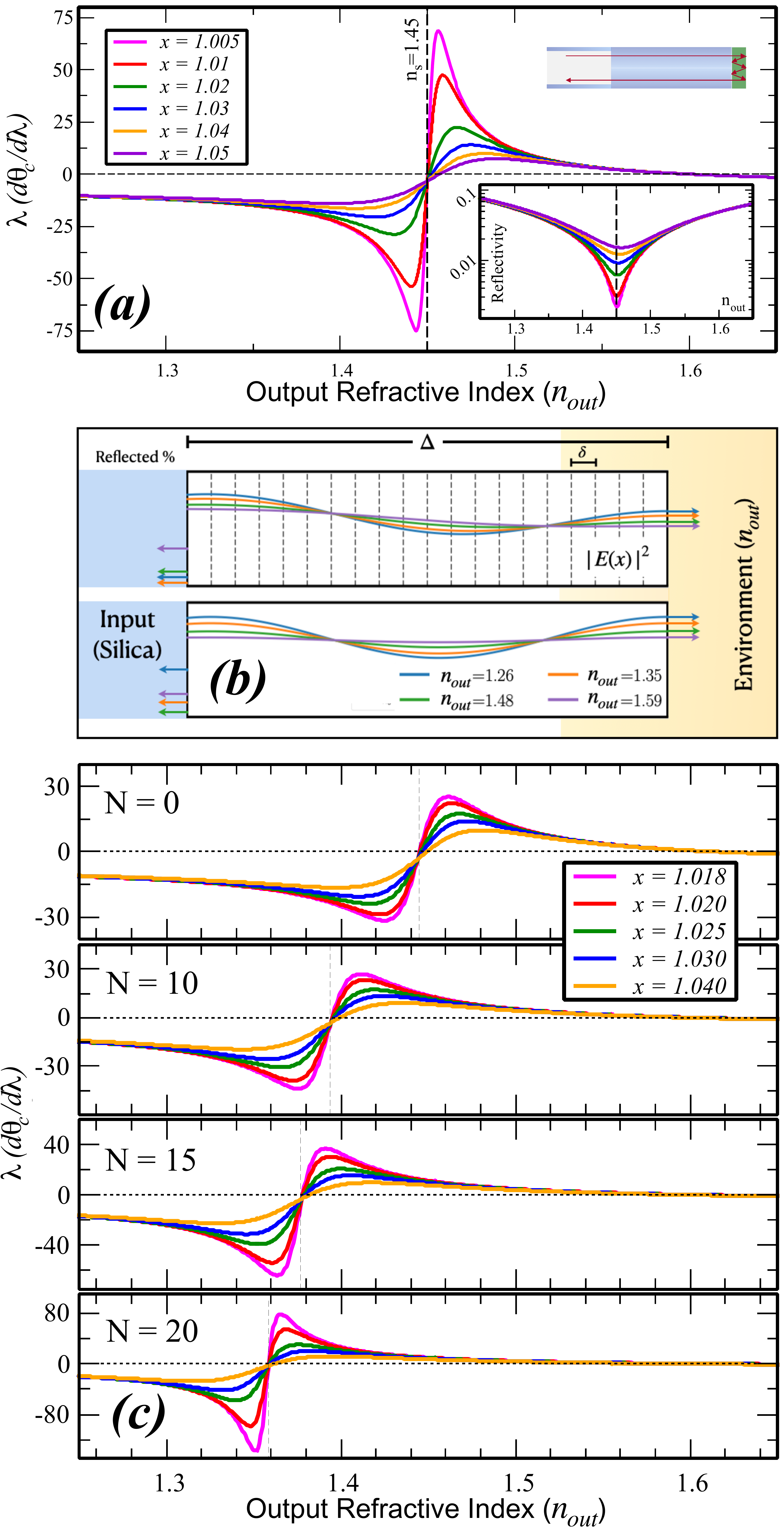}

\vspace{-0.4cm}

\caption{\,\label{fig:PlainResonance}\textit{(a)} Simulated plots of $\lambda d\theta_{\!\text{c}}\!/d\lambda$
(controlling the shift of the reflected fringes) and reflectivity
(in the inset), as a function of the output refractive index ($n_{{\scriptscriptstyle \!\text{out}}}$)
for a \textit{purely dielectric} and \textit{nearly anti-reflecting}
coating of index $n_{{\scriptscriptstyle \!\text{pol}}}\!=\!1.615$
and total width $\Delta\!=\!x\lambda/2n_{{\scriptscriptstyle \!\text{pol}}}$.\textit{\,(b)}
Scheme of the deformed intensity profile within the coating induced
by the conducting membranes. Simulated data is shown for a plain dielectric
coating (bottom) and the same coating intersected by $15$ graphene
sheets (top). To aid visualization, the vertical axis is to be taken
as proportional to the actual reflected intensity.\textit{ (c)} Simulated
environmental dependence of $\lambda d\theta_{\!\text{c}}\!/d\lambda$
for hybrid coatings with $N\!\!=\!0$\,-$20$ graphene sheets, operating
at a central wavelength of $\lambda_{\text{c}}\!=\!1550\text{nm}$.
The \textit{resonant response} is shifted to lower $n_{{\scriptscriptstyle \!\text{out}}}$,
as $N$ is increased. (color online)}

\vspace{-0.7cm}
\end{figure}

\noindent From Eq.\,(\ref{eq:ReflectionCoeff}), one extracts the
reflectivity and reflection phase shift as functions of the incident
wavelength. For our purposes, what really matters is to analyze the
variation of these quantities with $n_{\text{out}}$, provided the
coating is nearly anti-reflecting (i.e. $\Delta\!\approx\!n_{\text{pol}}\lambda/2$).
In Fig.\,\ref{fig:PlainResonance}a, we present such an analysis
for a dielectric coating ($n_{\text{pol}}\!=\!1.6$), near anti-reflecting
conditions, i.e. with the parameter $x\!=\!2\Delta n_{\text{pol}}/\lambda$
valued close to $1$. The main feature to be highlighted happens for
an index $n_{\text{out}}^{*}$, for which the absolute reflectivity
of the output surface vanishes and is accompanied by an abrupt (\textit{``resonant-like''})
behavior of $\left.\nicefrac{d\theta_{c}^{\lambda}}{d\lambda}\right|_{\lambda_{\text{c}}}\!\!$.
In the absence of conducting membranes, this behavior appears around
$n_{\text{out}}^{*}\!\!=\!n_{\text{s}}$ which signals that, even
for a coating with precise dimensions for destructive interference
of the backscattered waves\textit{ (i.e.} $x\!=\!1$), the interference
is only partial unless the reflectivity is similar on both ends of
the coating. This implies opposite discontinuities of the refractive
index in the two interfaces. If the dimensions of the coating slightly
depart from this $\lambda/2$-condition (or $x\!\approx\!1$) then
a nonzero reflectivity is obtained, accompanied by a broadening of
the \textit{``resonant-like''} response of $\left.\nicefrac{d\theta_{c}^{\lambda}}{d\lambda}\right|_{\lambda_{\text{c}}}\!\!$
to $n_{\text{out}}$ (as depicted in Fig.\,\ref{fig:PlainResonance}a).

The \textit{``resonant-like''} behavior described above is of the
utmost importance for this work. In fact, Eqs.\,(\ref{eq:lambda_eff})
directly imply that such a strong dependence of $\left.\nicefrac{d\theta_{c}^{\lambda}}{d\lambda}\right|_{\lambda_{\text{c}}}\!\!$
on $n_{\text{out}}$ leads to an enhanced sensitivity of the periods
in the interference fringes, to the environment in which the Fabry-Perot
microcavity is placed. In passing, it is also relevant to note that
$\left.\nicefrac{d\theta_{c}^{\lambda}}{d\lambda}\right|_{\lambda_{\text{c}}}\!\!$
increases with $n_{\text{out}}$ in this regime, meaning that the
periods of the interference fringes would have a negative environmental
response. However, as the closing of the microcavity is always silica
($n_{\text{s}}\!\approx\!1.45$), any purely dielectric coating would
only enhance its environmental sensitivity for refractive indices
$n_{\text{out}}^{*}\!\approx\!1.45$, which are typical of transparent
solid media.\,In order to operate this sensor on fluid environments,
one must be able to shift this \textsl{``resonant-like}'' behavior
towards lower values of $n_{\text{out}}^{*}$.\,As will be demonstrated,
this is the advantage of using a hybrid coating with embedded conducting
membranes, instead of a dielectric one. 
\begin{figure}[t]
\begin{centering}
\hspace{-0.1cm}\includegraphics[scale=0.22]{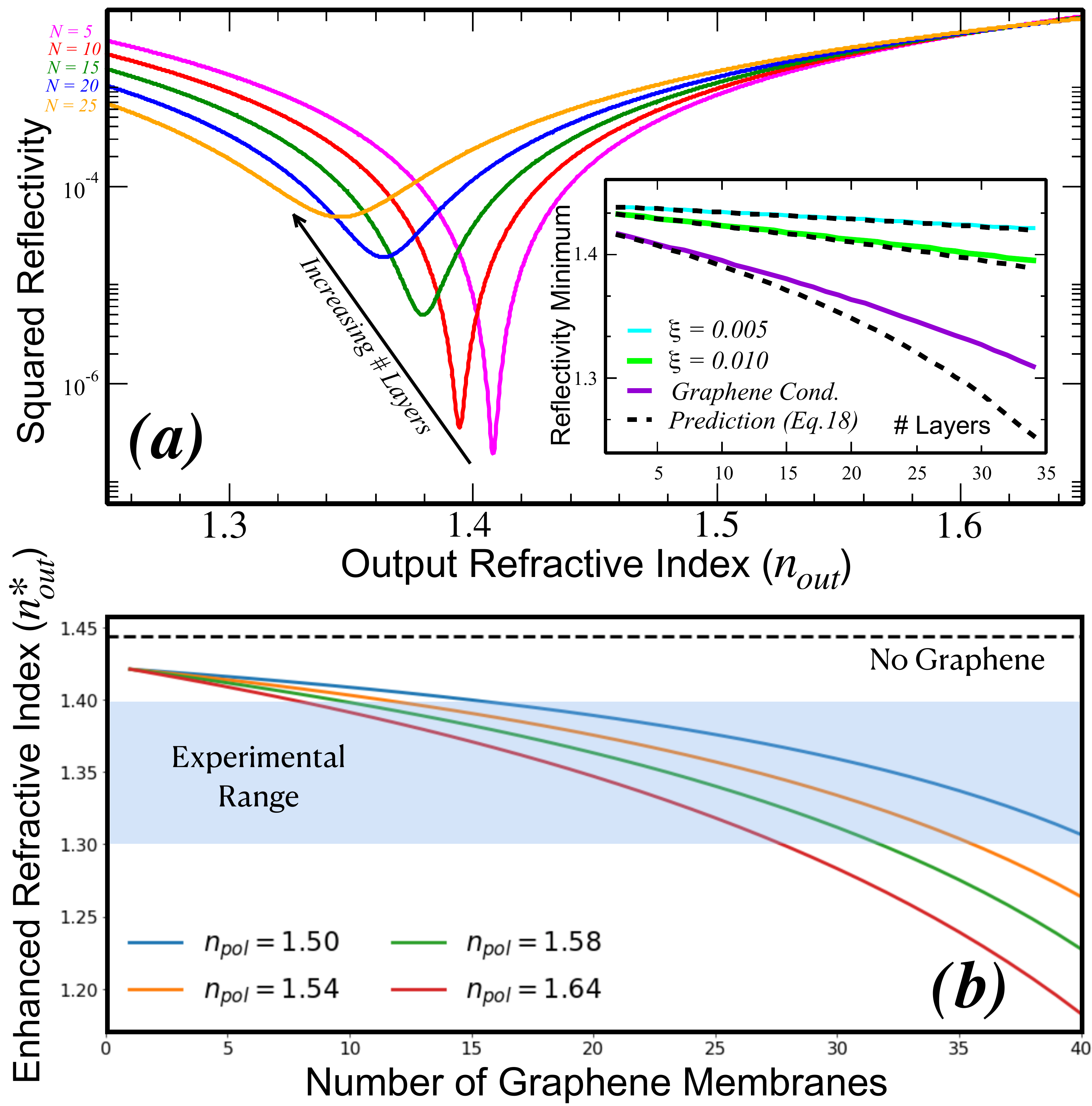}
\par\end{centering}
\vspace{-0.4cm}

\caption{\label{fig:CalibrationShift}\textit{(a)} Plots of the reflectivity
minima for simulated coatings having an increasing number of conducting
membranes. The position of the minimum coincides with the index $n_{{\scriptscriptstyle \text{\!out}}}$
around which the sensor's response to environmental changes is enhanced.
\textit{(b)} Examples of calibration curves that show the number of
graphene-dielectric bilayers needed to obtained an enhanced environmental
response around a particular refractive index. The different curves
refer to using different dielectric matrices in the coating and relevant
range to study typical liquid media is shaded in blue. (color online)}

\vspace{-0.4cm}
\end{figure}

Inside a hybrid coating there are conducting membranes that act as
segmentations of the bulk dielectric matrix in which they lie embedded.
Effectively, these membranes split the dielectric coating into chunks
of average width $\delta$, performing as conducting interfaces. The
presence of these interfaces greatly affect the profile of a wave
traveling inside, as they introduce additional internal reflections
and also localized dissipation of field energy. The TMM devised in
Sec.\,\ref{sec:TransferMatrix} takes all these effects into account
and, as shown in Fig.\,\ref{fig:PlainResonance}b, they lead to a
non-homogeneous field intensity that decreases along the width of
the coating. This deformation of the field profile leads to a change
in the condition for the absence of a reflected wave, thus requiring
a larger discontinuity in the refractive index at the output boundary.
As shown in Fig.\,\ref{fig:PlainResonance}c, the main consequence
is a downshift of $n_{\text{out}}^{*}$ which now enhances the environmental
sensitivity at refractive indices within the range of $1.3\!-\!1.4$.

Before moving on to the experimental proof-of-concept, it is important
to recognize that the shift in $n_{\text{out}}^{*}$ depends on the
parameters of the coating. Namely, we show in Fig.\,\ref{fig:PlainResonance}c
that $n_{\text{out}}^{*}$ decreases monotonically with the number
of conducting membranes, provided the full width of the coating is
kept roughly the same. Instead of looking at the resonance, the changes
in $n_{\text{out}}^{*}$ can be conveniently and fully characterized
by tracking the reflectivity minimum. According to our previous interpretation
of this \textit{``resonance-like''} behavior, a minimum in the reflectivity
must always accompany the resonance in $\left.\nicefrac{d\theta_{c}^{\lambda}}{d\lambda}\right|_{\lambda_{\text{c}}}\!\!$.
This analysis is done in Fig.\,\ref{fig:CalibrationShift}a and,
for low enough conductivity of the membranes, the minima depend on
the coating parameters as,
\begin{equation}
n_{\text{out}}^{*}\!\!=\!\sqrt{\frac{n_{\text{s}}^{2}-\mu_{0}c\sigma_{\lambda}n_{\text{s}}(N\!+\!1)}{1-\mu_{0}c\sigma_{\lambda}n_{\text{s}}(N\!-\!1)/n_{\text{pol}}^{2}}},
\end{equation}
where $n_{\text{s}}$ is the refractive index of silica, $\mu_{0}$
the vaccum magnetic permeability, $c$ the speed of light and $N$
the total number of conducting membranes. Finally, this expression
can be used to create approximate calibration curves (see Fig.\,\ref{fig:CalibrationShift}b)
that serve to guide the engineering of a coating configuration that
is appropriate to enhance the environmental sensitivity around a pre-established
value of the output refractive index.

\vspace{-0.3cm}

\section{\label{sec:RExperimental}Experimental Demonstration}

\vspace{-0.1cm}

The theoretical proposal made in previous sections is now put to the
test on experimental grounds. Our prototypical device consists of
a single-mode optical fiber (SMF) terminated by a cascaded Fabry-Perot
microcavity interferometer (FPMI). A photograph obtained from an optical
microscope is shown in Fig.\,\ref{fig:Experimental-setup-composed}.
In order to assess the influence of an hybrid coating on the environmental
sensitivity of its interference pattern, we perform a comparative
study using two sensing devices: i) an FPMI sensor coated by a single
layer of bulk dielectric polymer and ii) a similar sensor having a
multilayered PEI/GO hybrid coating.

In the following, we begin by detailing technical aspects of the device's
production method and measurement techniques. Afterwards, we present
experimental measurements and compare them to the predictions of our
theoretical framework.

\vspace{-0.5cm}

\subsubsection{Fabrication of the Fabry-Perot Microcavity Interferometer}

\vspace{-0.2cm}

To build the FPMI, a silica capillary tube with an internal (external)
diameter of 75\textmu m (125\textmu m) was fusion spliced (Sumitomo
Type-72C) to the SMF's termination and then cleaved to the desired
length ($D$), using a fiber cleaver (Sumitomo FC-6RS). To close the
FPMI, a second SMF was fusion spliced to the free end face of the
capillary and then cleaved to a length $d$. The splicing processes
were all made by using electric arcs centered far away from the capillary
region. This way, we could avoid its collapse and thus keep straight
enough sides in order to minimize transverse optical losses.

\vspace{-0.5cm}

\subsubsection{Production of the PEI and PEI/GO coatings}

\vspace{-0.2cm}

For producing the optical coatings, we employed polyethylenimine (PEI)
as our dielectric medium ($n_{\!\text{out}}\!=\!1.615$). This polymer
was acquired from Sigma-Aldrich (catalog number P3143) as a water-based
solution with a $50\%$ concentration ($w/v$). The PEI solution was
further diluted using ultra-pure water (Milli-Q water), achieving
a lower concentration of $2\!\times\!10^{-2}\text{\!}M$. For producing
the multilayered hybrid coatings, we used graphene-oxide obtained
from a commercial GO solution (Sigma-Aldrich, catalog number 777676)
with a concentration of 4\,mg/mL. The latter was then diluted in
ultra-pure water to match the same concentration of the diluted PEI
solution.

All FPMIs used in this study were coated using a layer-by-layer dip
coating method. For purely dielectric (PEI) coatings, the fiber was
dipped into the diluted PEI solution for 1\,min, followed by a brief
rinse in ultra-pure water to remove non-adsorbed residues. This process
was repeated 10 times, in order to reach a significant film thickness.
In the case of the PEI/GO coatings, a similar procedure was followed,
with the FPMIs being dipped alternately into the PEI and the GO solution,
with a rinse and drying step in between. Here the repetition went
until the desired number of bilayers ($N\!=\!10$) was attained.

\begin{figure}[t]
\includegraphics[scale=0.17]{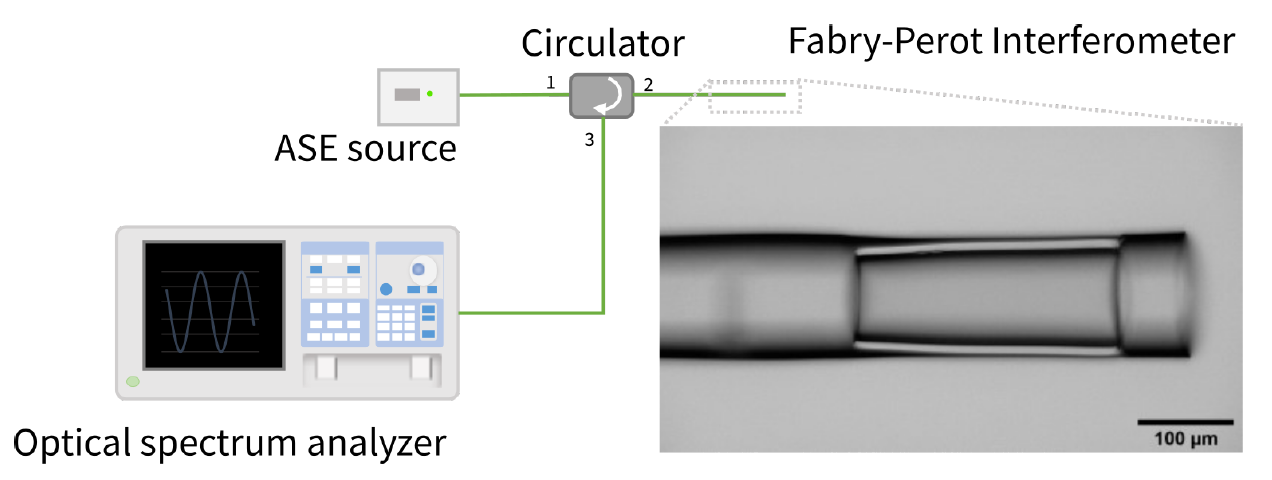}

\vspace{-0.4cm}

\caption{\label{fig:Experimental-setup-composed}Experimental setup composed
by an amplified spontaneous emission (ASE) broadband source, an optical
circulator, and an optical spectral analyzer (OSA). Microscope photograph
of the fabricated Fabry-Perot interferometer, with an air cavity length
of 258\,\textmu m and a silica slab length of 64\,\textmu m. (color
online)}

\vspace{-0.6cm}

\end{figure}

\vspace{-0.5cm}

\subsubsection{Optical Measurement Setup}

\vspace{-0.2cm}

The experimental setup used for the characterization of the FPMIs
is schematically depicted in Fig.\,\ref{fig:Experimental-setup-composed}.
It consisted of an amplified spontaneous emission (ASE) broadband
source centered at $\lambda_{\text{c}}\!=\!1550\text{nm}$ with a
bandwidth of (approximately) 100\,nm, an optical circulator, and
an optical spectrum analyzer (OSA - Yokogawa AQ6370C).

\vspace{-0.3cm}

\subsection{Experimental Results}

\vspace{-0.2cm}

\begin{figure}[t]
\vspace{-0.3cm}

\includegraphics[scale=0.27]{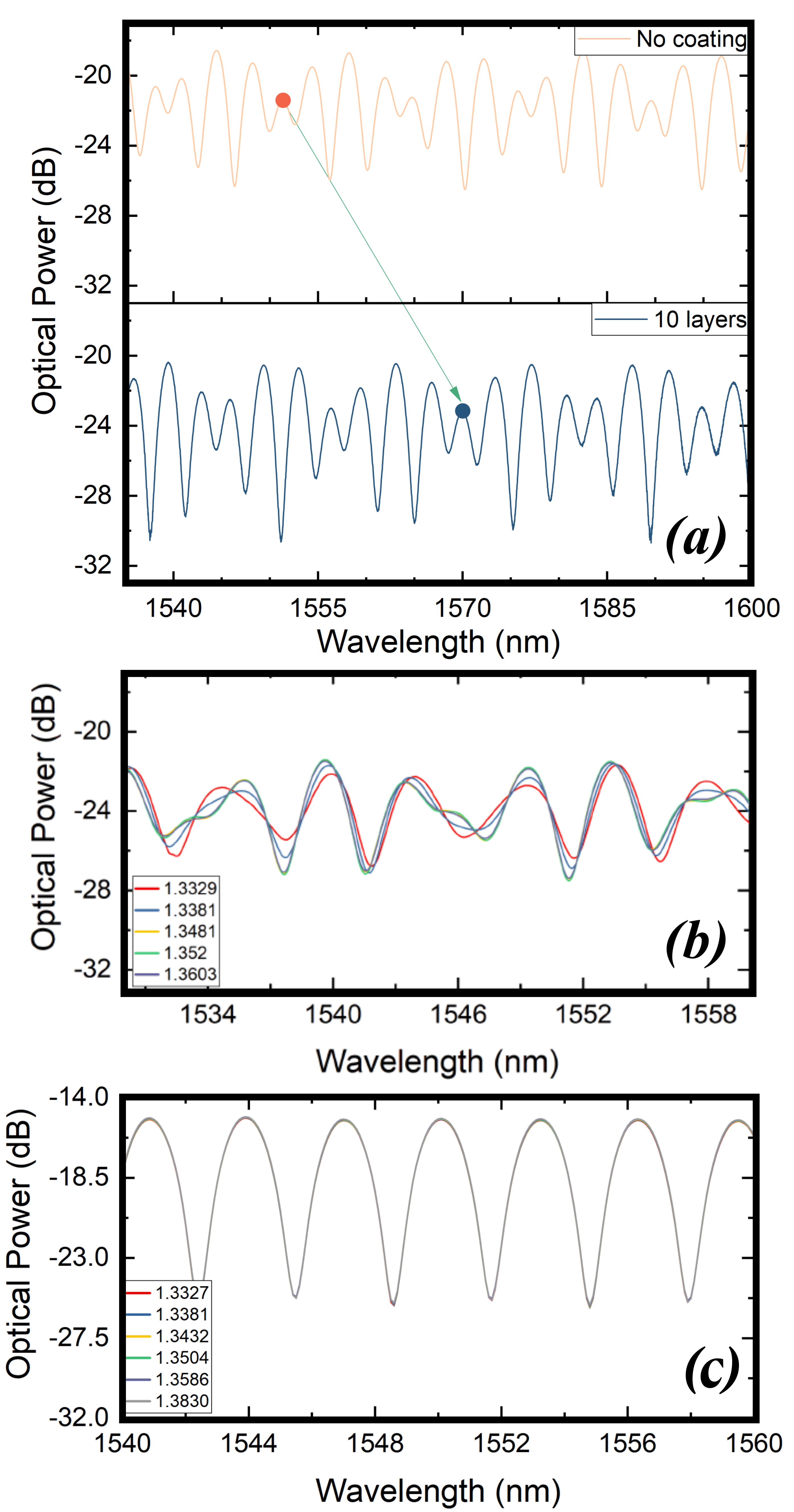}

\vspace{-0.4cm}

\caption{\label{fig:ExperimentalResults}\textit{(a)} Measured reflected spectrum
before and after the application of a coating with 10 PEI/GO bilayers.
The arrow makes evident the overall shift of the interference pattern
(in wavelength) caused by introducing the output coating. \textit{(b)}
Reflected spectrum from the FPMI coated by 10 PEI/GO, measured with
the cavity immersed by liquid environments having different refractive
indices ($n_{{\scriptscriptstyle \!\text{out}}}\!\!=1.3329\!-\!1.3603$).
The sensitivity of the pattern is to be contrasted with \textit{(c)},
where an identical analysis was done using a cavity coated by PEI
with roughly the same width. All the curves appear superposed. (color
online)}
\vspace{-0.6cm}
\end{figure}
The influence of coating the output surface of an FPMI was studied
by comparing the reflected spectrum before and after the dip coating
process. In the presence of a PEI coating, no significant shifts or
intensity changes were observed in the reflected spectrum, as obtained
within an air environment. In contrast, the PEI/GO FPMI showed visible
changes in the interference fringes, which appeared shifted in wavelength,
with decreased intensity but a larger visibility. In Fig.\,\ref{fig:ExperimentalResults}a,
we plot the reflected signal measured before and after a sensor was
coated with 10 PEI/GO bilayers. The shift in the fringes is clearly
marked.

Besides comparing the operation of coated and uncoated FPMIs, we focus
our analysis on the influence of such coatings in their sensitivity
to environmental changes. Namely, we characterize the dependence of
the reflected pattern on the refractive index of the output medium
by immersing the sensors in calibrated liquids, with indices ranging
from $n_{\text{out}}\!=\!1.3327$ to $n_{\text{out}}\!=\!1.3830$\,\footnote{These indices were determined using an Abbe refractometer (Atago DR-A1),
with a light source with a wavelength approximately equal to sodium
D-line (589.3\,nm).}. The spectra of reflected signals obtained with a PEI/GO-coated sensor
are shown in Fig.\,\ref{fig:ExperimentalResults}b, where it is evident
that changing the output medium induces measurable variations in both
phase and intensity of the interferences fringes. No such response
was obtained using a PEI-coated FPMI, as shown in Fig.\,\ref{fig:ExperimentalResults}c,
which indicates an increased environmental sensitivity of the system
in the presence of a hybrid coating. 

\begin{figure}[t]
\hspace{-0.1cm}\includegraphics[scale=0.26]{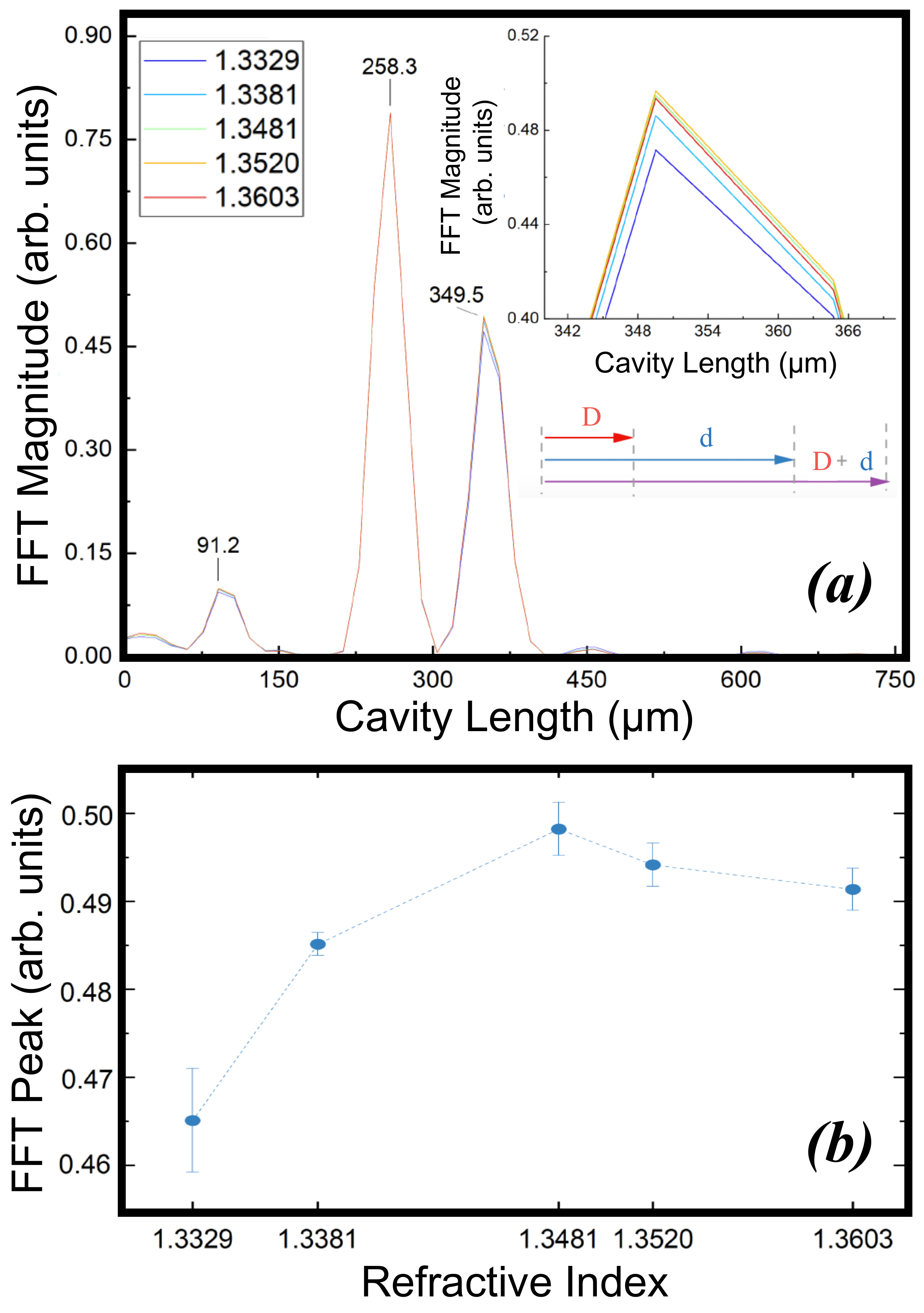}

\vspace{-0.4cm}

\caption{\label{fig:ExperimentalResults-1}\textit{(a)} Fourier transform (FFT)
of the reflected spectra measured with the sensor inside outer liquid
media with varying refractive indices, coated by 10 PEI/GO bilayers.
\textit{(b) }Study of the dependence of the intensity in third peak
on the output refractive index. The error bars are due to intrinsic
spectral resolution of the OSA which appear as random fluctuations
in measurements made over time, using the same device. (color online)}
\vspace{-0.7cm}
\end{figure}

The variations in the reflected signal of the PEI/GO sensor can be
more finely evaluated by performing a spectral decomposition\,\footnote{The FFT was calculated after converting the dB-reflected signal into
linear optical power (W).} through a fast Fourier transform (FFT). The result is shown in Fig.\,\ref{fig:ExperimentalResults-1}a,
where the FFT amplitudes appear plotted as functions of vaccum optical
cavity lengths. From our previous theoretical analysis, it is not
surprising that the FFT features exactly three peaks, associated to
the three Fourier components of Eq.\,(\ref{eq:InterferencePattern_3Waves-1}).
The first peak ($\Lambda_{1}$), at a cavity length of $91.2$\,\textmu m
(corresponding to a $d\!=$63\,\textmu m length for $n_{\text{s}}\!\!=\!\!1.4444$),
is related to the optical cavity composed by fused silica and PEI/GO
coating. The second peak ($\Lambda_{2}$), related to the air cavity,
is located at a cavity length of $D\!=\!258.3$\,\textmu m, while
the third peak ($\Lambda_{3}$) is related to the optical cavity composed
by air, silica slab, and the PEI/GO coating. Note that the estimated
cavity length values are in full accordance with the ones determined
by inspection of the microscope photograph in Fig.\,\ref{fig:Experimental-setup-composed}.
Furthermore, small intensity variations are observed in Fig.\,\ref{fig:ExperimentalResults-1}a
for the first and third peaks, while the air cavity peak remains unaffected
by changes in the output refractive index. These results are consistent
with the understanding that only $\Lambda_{1}$ and $\Lambda_{3}$
(as defined in Eq.\,(\ref{eq:InterferencePattern_3Waves-1})) can
be affected by the reflection coefficient of the output coating and,
thus being sensitive to the output medium. As an example, we also
plot the FFT peak intensity of $\Lambda_{3}$ as a function of $n_{\text{out}}$
in Fig.\,\ref{fig:ExperimentalResults-1}b.

\begin{figure}[b]
\vspace{-0.4cm}

\hspace{-0.15cm}\includegraphics[scale=0.16]{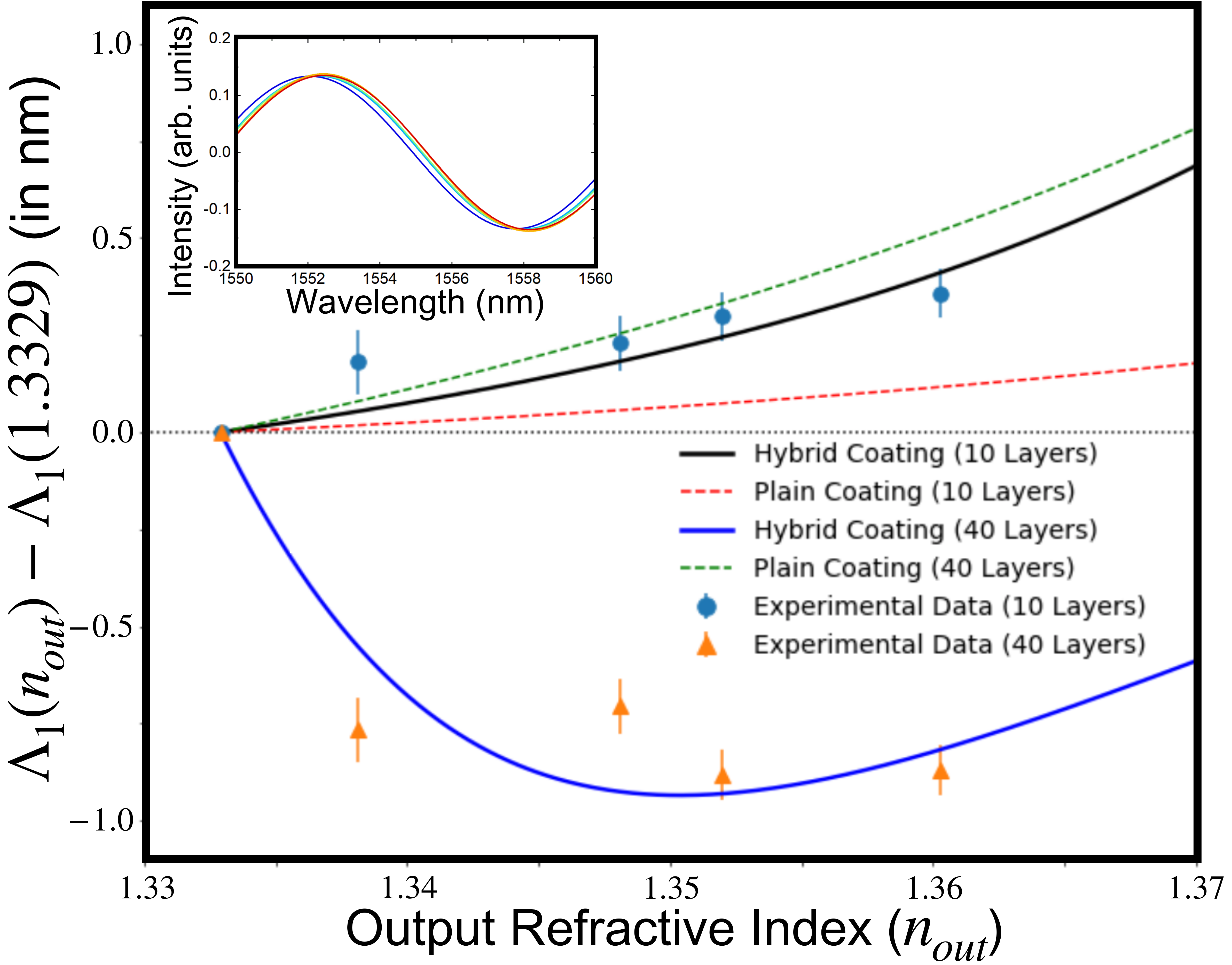}

\vspace{-0.4cm}

\caption{\label{fig:ExperimentalResults2}Fit of the spectral shifts in the
shortest period ($\Lambda_{1}$) measured from the filtered FFT of
the reflected fringes to the theoretical model. Two data sets are
shown for identical sensors coated by $10$ (blue points) and $40$
GO/PEI bilayers (orange points). The solid lines represent the corresponding
theoretical curves, obtained for reasonable values of the free parameters
in the model --- $\xi\!=\!0.0188$, $n_{\text{pol}}\!=\!1.617$ ---
and a distance between consecutive GO monolayers of $\delta\!=\!95.8$\,nm
and $\delta\!=\!95.6$\,nm, respectively. For a comparison, the corresponding
curves in the absence of GO membranes (i.e. $\xi\!=\!0$) are shown
as dashed lines. The inset represents the band-filtered signal reconstructed
by an inverse FFT for the coating of $10$ bilayers. The error bars
are due to intrinsic spectral resolution of the OSA which appear as
fluctuations in measurements made over time. (color online)}

\vspace{-0.4cm}
\end{figure}

The changes in the shape of the interference pattern depicted in Fig.\,\ref{fig:ExperimentalResults}b,
point towards an environmentally-induced variation not only of the
FFT peak intensities, but also of their locations. Such changes in
the periodicity of the pattern are consistent with our earlier theoretical
predictions and therefore provide the sought after connection between
the experiments and theory presented here. In order to analyze these
shifts in the Fourier components, we begin by isolating one of the
sensitive peaks using a band-pass filter in the FFT. For concreteness,
we isolate the first peak ($\Lambda_{1}$) and obtain the filtered
spectrum presented in the inset of Fig.\,\ref{fig:ExperimentalResults2}.
The filtered signal now exhibits an evident phase variation which
actually amounts to a shift of the period $\Lambda_{1}$ induced by
an increase of the output refractive index. In Fig.\,\ref{fig:ExperimentalResults2},
the shifts in $\Lambda_{1}$ are represented against our theoretical
predictions for two FPMI sensors coated with $N\!=\!10$ and $N\!=\!40$
conducting membranes (GO) embedded into a dielectric (PEI) matrix
($n_{\text{pol}}\!=\!1.617$\,\citet{PEI_Web}). Surprisingly, by
using the simplified model of Sec.\,\ref{sec:TransferMatrix} and
a real optical conductivity slightly smaller than pristine graphene
($\xi\!=\!0.0188$\,\footnote{As opposed to $\xi\!\approx\!0.0229$ for pristine graphene sheets
working in the mid- to near-infrared regime.}), the agreement between experimental data and the theoretical curves
is reasonably good, for both cases. The average distance between consecutive
GO sheets was slightly adjusted for the fit ($\delta\!=\!95.8$\,nm
and $\delta\!=\!95.6$\,nm, respectively) but the values are close
enough to indicate a good consistence in the production of the multilayered
coating. Given the imperfections expected to be present in a GO/PEI
coating produced using a chemical layer-by-layer technique, we can
argue our experimental results to be consistent with the downshift
of a \textit{``resonant-like''} behavior in the reflection phase
shift at the output surface of the FPMI, as theoretically predicted
for hybrid coatings in Sec.\,\ref{sec:Resonance}.

\vspace{-0.4cm}

\section{\label{sec:Conclusions}Conclusions and Outlook}

\vspace{-0.2cm}

We have proposed and experimentally demonstrated an interferometric
sensor, based on a Fabry-Perot microcavity built at the end of a single-mode
optical fiber and covered by a thin multilayered hybrid coating made
of two-dimensional (graphene-like) conducting membranes embedded within
a dielectric matrix. Unlike more conventional Fabry-Perot sensors,
the proposed device is able to sense changes in the refractive index
of the output medium by translating them to measurable shifts in the
Fourier periods that compose the reflected spectrum. Our theoretical
study has shown that the aforementioned changes in the interference
pattern can be attributed to an enhanced environmental sensitivity
of the reflection phase shift introduced by the output coating. The
latter is caused by complex internal reflection, interference and
energy-loss processes among the different conducting surfaces contained
within the coating. Moreover, as the number of layers in the coating
increases, the sensitivity is enhanced for lower values of the output
refractive index, which allows the engineering of a sensor that works
within range of indices typically found on liquid media ($n_{\text{out}}\!\approx1.3\!-\!1.4\!$).
The theoretical predictions were further tested against reflected
spectral measurements done on an experimental realization of this
device that employed GO/PEI multilayered structures produced using
a cost effective layer-by-layer chemical dip coating process. Despite
inherent imperfections on the coating's structure and the possible
influence of other external physical parameters (e.g. thermal sensitivity),
a sensible agreement was obtained between measurements and theoretical
predictions. Besides demonstrating the concept, this agreement further
encourages investment on these types of devices for future real-life
sensing applications.

\begin{figure}[t]
\vspace{-0.7cm}
\begin{centering}
\includegraphics[scale=0.1]{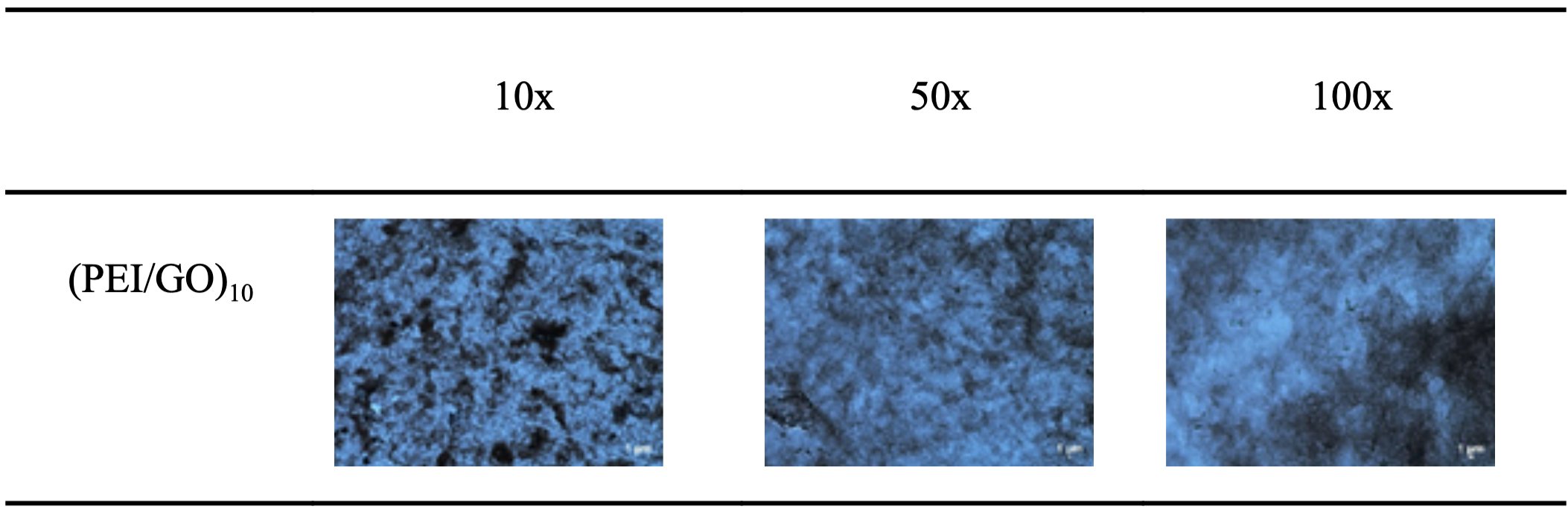}
\par\end{centering}
\vspace{-0.3cm}

\caption{\label{fig:Optical-microscopy-images}Optical microscopy images of
a film of 10 PEI/GO bilayers deposited on glass. The roughness of
the polymeric film is clear from the picture.}

\vspace{-0.6cm}
\end{figure}

As a closing remark, we comment on the technological applicability
of the sensing devices described in this work. In reality, for these
hybrid coatings to be of practical use, a more consistent production
method that creates more resilient/durable hybrid coatings is imperative.
In particular, the interference phenomenon described here is very
sensitive to imperfections in the layered structure, namely the film
roughness shown in Fig.\,\ref{fig:Optical-microscopy-images}. This
roughness is a typical feature known to occur in polymeric films produced
using layer-by-layer techniques\,\citep{Ferreira2007,Duarte2013,raposo2015}
and can eventually be controlled. Surely, the use of vapor phase deposition
techniques could potentially allow more perfect structured coatings
to be made. However, such methods would be highly inefficient, and
likely more expensive, ways to produce these multilayered coatings.
Nevertheless, the basic concepts underlying the operation of these
hybrid coatings are arguably more general than our current experimental
realization and this work is to serve as a spark for future technological
developments.

\vspace{-0.5cm}
\begin{acknowledgments}
The authors acknowledge financial support by the Portuguese Foundation
for Science and Technology (FCT) within the Strategic Funding UIDB/04650/2020
and COMPETE 2020 program in FEDER component (European Union), through
Projects No.\,POCI-01-0145-FEDER-028887\,\,and UltraGraf (M-ERA-NET2/0002/2016).
J.P.S.P. and C.S.M were further supported by FCT Ph.D. grants PD/BD/142774/2018
and SFRH/BD/135820 /2018, respectively. J.P.S.P. thanks S.\,M. João,
C.\,D. Fernandes, Prof.\,\,J.\,M.\,B.\,Lopes dos Santos and
Prof.\,\,N.\,M.\,R.\,Peres for profitable discussions and useful
comments on this work.
\end{acknowledgments}

\appendix

\vspace{-0.3cm}

\section{\label{BoundaryConditions}Electrodynamic boundary conditions in
conducting interfaces}

\vspace{-0.3cm}

\noindent The existence of sharp interfaces separating different dielectric
media are known to introduce discontinuities on components of electric
and/or magnetic fields. In the simplest case, these interfaces are
electrically neutral and cannot support charge currents. For this
work, however, we must consider cases in which this last condition
does not hold, because the interfaces are two-dimensional conducting
membranes (e.g. graphene monolayers). In this Appendix, we will derive
the general boundary conditions for such a planar interface, assuming
that the two dielectrics have refractive indices $n_{1}$ and $n_{2}$,
while the interface is a linear conducting membrane described by a
complex conductivity $\sigma_{\lambda}\!=\!\sigma'_{\!\lambda}\!+\!i\sigma''_{\!\lambda}$. 

To obtain these boundary conditions, we begin by considering the interface
to be planar. Despite not involving loss of generality, this geometrical
arrangement allows the construction of the usual pillbox gauss surface
(Fig.\,\ref{fig:BoundaryConditions}a) and Ampère square circuit
(Fig.\,\ref{fig:BoundaryConditions}b) that allows us to apply Maxwell's
equations in integral form:
\begin{figure}[t]
\vspace{-0.4cm}
\begin{centering}
\includegraphics[scale=0.13]{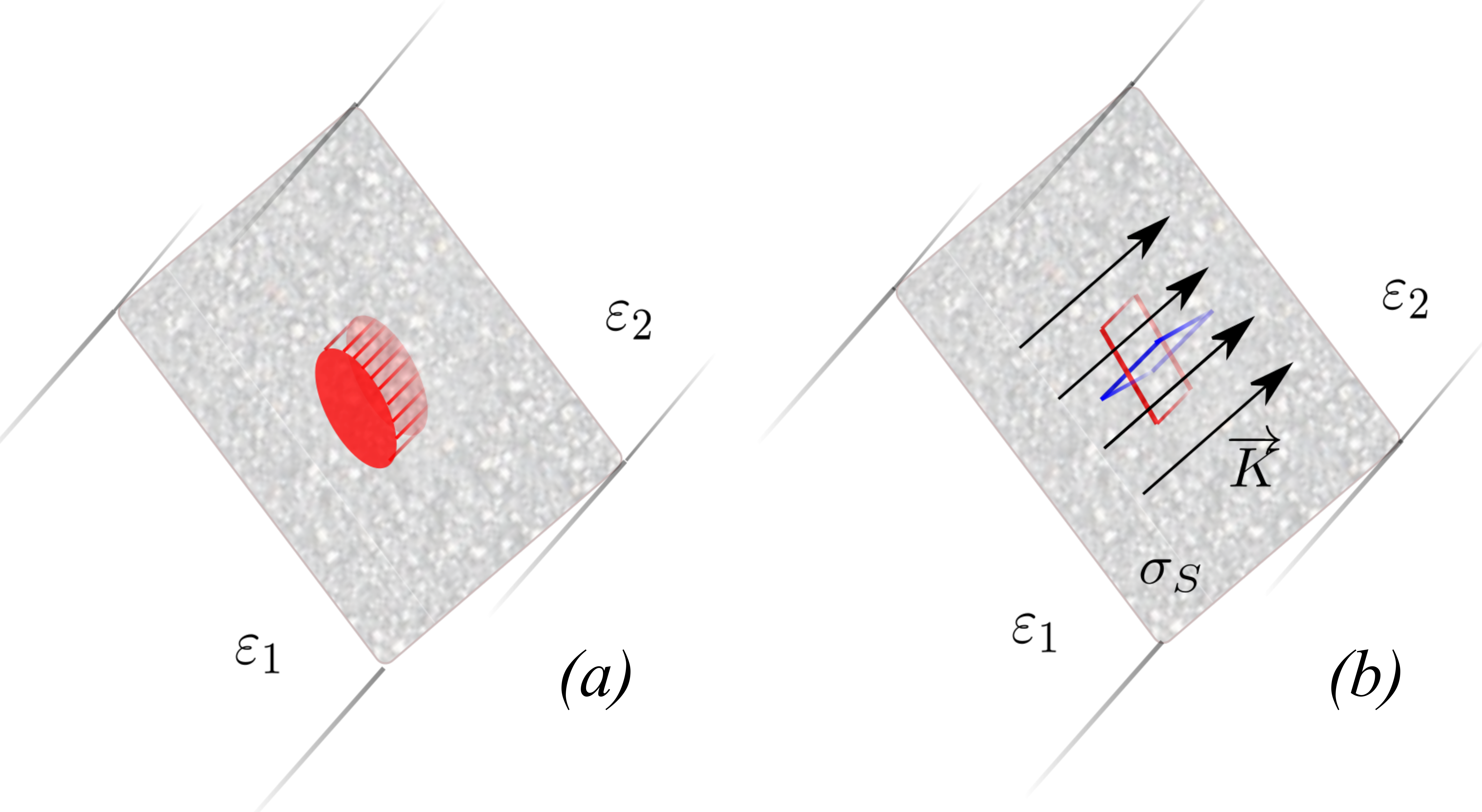}
\par\end{centering}
\vspace{-0.4cm}

\caption{\label{fig:BoundaryConditions}(a) Pillbox Surface traversing the
interface between the two dielectric media. (b) Square circuit used
to derive the boundary conditions for the parallel components. (color
online)}

\vspace{-0.7cm}
\end{figure}

\vspace{-0.3cm}

\begin{subequations}
\begin{equation}
\varoiint_{S_{\mathcal{V}}}\!\!\!\!\mathbf{D}\!\left(\mathbf{r},t\right)\!\cdot\!\mathbf{dS}\!=\!\!\iiint_{\mathcal{V}}\!\!\!\rho\!\left(\mathbf{r},t\right)d\mathbf{r}\label{eq:Maxwell1}
\end{equation}

\vspace{-0.6cm}

\begin{equation}
\varoiint_{S_{\mathcal{V}}}\!\!\!\!\mathbf{B}\!\left(\mathbf{r},t\right)\!\cdot\!\mathbf{dS}\!=\!0\label{eq:Maxwell2}
\end{equation}

\vspace{-0.6cm}

\begin{equation}
\ointclockwise_{\mathcal{C}}\!\!\mathbf{E}\!\left(\mathbf{r},t\right)\!\cdot\!\mathbf{dS}\!=\!-\varoiint_{S_{\mathcal{C}}}\!\!\left[\frac{\partial}{\partial t}\mathbf{B}\!\left(\mathbf{r},t\right)\right]\!\cdot\!\mathbf{dS}\label{eq:Maxwell3}
\end{equation}

\vspace{-0.6cm}

\begin{align}
\ointclockwise_{\mathcal{C}}\!\!\mathbf{B}\!\left(\mathbf{r},t\right)\!\cdot\!\mathbf{dS}\! & =\!\mu_{0}\varoiint_{S_{\mathcal{C}}}\!\!\!\mathbf{J}\!\left(\mathbf{r},t\right)\!\cdot\!\mathbf{dS}\label{eq:Maxwell4}\\
 & \quad-\mu_{0}\varoiint_{S_{\mathcal{C}}}\!\!\left[\frac{\partial}{\partial t}\mathbf{D}\!\left(\mathbf{r},t\right)\right]\!\cdot\!\mathbf{dS},\nonumber 
\end{align}
\end{subequations}
where $\mathbf{E}$, $\mathbf{B}$ and $\mathbf{D}$ are the electric
magnetic and electric displacement fields, respectively. In order
to keep the discussion as focused as possible, we will assume that
all these fields are harmonic in space-time, e.g. $\mathbf{E}(\mathbf{r},t)\!=\!\mathbf{E}_{0}\exp\left(i\mathbf{k}\!\cdot\!\mathbf{r}\!-\!i\omega t\right)$,
and that the interface is aligned with the plane $z\!=\!0$. Assuming
an wavelength $\lambda\!=\!2\pi/\abs{\mathbf{k}}$, the ohmic surface
currents at the interfaces are simply 

\vspace{-0.4cm}
\begin{equation}
\mathbf{K}\!(\mathbf{r},t)\!=\!\mathbf{K}_{0}\exp\left(i\mathbf{k}^{\parallel}\!\cdot\!\mathbf{r}\!-\!i\omega t\right),\label{eq:Currents}
\end{equation}
where $\mathbf{K}_{0}\!=\!\sigma_{\lambda}\!\mathbf{E}_{0}^{\parallel}$.
Hereafter, we will use $\parallel$ ($\perp$) to represent vectors
which are parallel (perpendicular) to the interface. Since the current
of Eq.\,(\ref{eq:Currents}) is time-dependent, the condition of
local current conservation implies that a time-dependent change density
wave is also usually generated. The latter can be obtained from the
two-dimensional continuity equation, 

\vspace{-0.5cm}

\begin{equation}
\frac{\partial\rho_{\text{s}}(\mathbf{r},t)}{\partial t}=\nabla_{xy}\mathbf{K}\left(\mathbf{r},t\right),\label{eq:Continuity}
\end{equation}
which yields $\rho_{\text{s}}\!(\mathbf{r},t)\!=\!\rho_{\text{s}}^{0}\exp\left(i\mathbf{k}^{\perp}\!\cdot\!\mathbf{r}\!-\!i\omega t\right)$,
where 

\vspace{-0.4cm}
\begin{equation}
\rho_{\text{s}}^{0}\!=\!\frac{\sigma_{\lambda}}{\omega}\mathbf{E}_{0}^{\parallel}\!\cdot\!\mathbf{k}.\label{eq:ChargeDensityWave}
\end{equation}
If we now use the surface current and charge densities of Eqs.\,(\ref{eq:Currents})
and (\ref{eq:ChargeDensityWave}) and apply the integral Maxwell's
equations to the constructions of Fig.\,\ref{fig:BoundaryConditions},
we obtain the following general conditions for the fields in the two
sides of the planar interface:

\vspace{-0.4cm}

\begin{subequations}
\begin{align}
\begin{array}{cc}
\mathbf{E}_{0,1}^{\parallel} & \!=\!\mathbf{E}_{0,2}^{\parallel}\end{array}
\end{align}

\vspace{-0.5cm}

\begin{equation}
\varepsilon_{1}\mathbf{E}_{0,1}^{\perp}\!=\!\varepsilon_{2}\mathbf{E}_{2}^{\perp}\!+\!\frac{\sigma_{\lambda}}{\omega}\mathbf{E}_{0,1}^{\parallel}\!\cdot\!\mathbf{k}\label{eq:SurfaceCurrent}
\end{equation}

\vspace{-0.5cm}

\begin{align}
\begin{array}{cc}
\mathbf{B}_{0,1}^{\perp} & \!=\!\mathbf{B}_{0,2}^{\perp}\end{array}
\end{align}

\vspace{-0.5cm}

\begin{equation}
\mathbf{B}_{1}^{\parallel}\!=\!\mathbf{B}_{2}^{\parallel}+\mu_{0}\sigma_{\lambda}\!\mathbf{\hat{n}}\!\times\!\mathbf{E}_{0,1}^{\parallel},\label{eq:B_Disc}
\end{equation}
\end{subequations}
where $\varepsilon_{1}$ are the dielectric constants of the two media.
Finally, if we consider the case of perpendicular incidence, then
the perpendicular components are zero and $\mathbf{\hat{n}}\!\times\!\mathbf{E}_{0,1}^{\parallel}\!=\!\mathbf{E}_{0,1}$.
This way, we recover the special case quoted in Eqs.\,(\ref{eq:ElectricFieldContinuity})-(\ref{eq:MagneticFieldCondition}). 

\bibliography{References}

\end{document}